\documentclass[prb, aps, twocolumn, 10pt, floatfix,showpacs,showkeys, superscriptaddress]{revtex4-2}

\usepackage{graphicx}
\usepackage{epstopdf}
\usepackage{amsmath}

\usepackage{color}

\usepackage[utf8]{inputenc}
\usepackage{xcolor}

\begin{document}

\title{Coexistence of charge density wave and field-tuned magnetic states in TmNiC$_2$}

\author{Kamil K. Kolincio}

\affiliation{Faculty of Applied Physics and Mathematics, Gdansk University of Technology,
Narutowicza 11/12, 80-233 Gdansk, Poland}
\email{corresponding author: kamil.kolincio@pg.edu.pl}

\author{Marta Roman}

\affiliation{Institute of Solid State Physics, TU Wien, Wiedner Hauptstrasse 8–10, A-1040 Wien, Austria}

\affiliation{Institute of Physics and Applied Computer Science, Faculty of Applied Physics and Mathematics, Gdansk University of Technology, Narutowicza 11/12, 80-233 Gdansk, Poland}

\author{Valerio Tammurello}
\affiliation{Dipartimento di Fisica, Sapienza University of Rome, Piazzale Aldo Moro 5, 00185 Rome, Italy}

\author{Simone Di Cataldo}
\affiliation{Dipartimento di Fisica, Sapienza University of Rome, Piazzale Aldo Moro 5, 00185 Rome, Italy}

\author{Daniel Matulka}
\affiliation{X-Ray Center, TU Wien, Getreidemarkt 9, A-1060 Wien, Austria}

\author{Sonia Francoual}
\affiliation{Deutsches Elektronen-Synchrotron DESY, 22607 Hamburg, Germany}

\author{Berthold St\"{o}ger}
\affiliation{X-Ray Center, TU Wien, Getreidemarkt 9, A-1060 Wien, Austria}

\author{Herwig Michor} 
\affiliation{Institute of Solid State Physics, TU Wien, Wiedner Hauptstrasse 8–10, A-1040 Wien, Austria}

\begin{abstract}
Exploring the relations between coexisting, cooperative, or competing types of ordering is a key to identify and harness the mechanisms governing the mutual interactions between them, and to utilize their combined properties. We have experimentally explored the response of the charge density wave (CDW) to various antiferromagnetic, metamagnetic, and field-aligned ferromagnetic states that constitute the magnetic phase diagram of TmNiC$_2$. The high resolution x-ray diffraction experiment employing synchrotron radiation at low temperature and high magnetic field, allowed to follow the superstructure satellite reflections, being a sensitive probe of CDW.
This investigation not only reveals direct evidence that the charge density wave avoids even a partial suppression in the antiferromagnetic ground state but also proves that this state coexists, without any visible signatures of weakening, in the entire dome of the magnetically ordered phases, including the field-aligned ferromagnetic state.  The calculations of the electronic and phonon structures support the experiment, revealing that the dominant contribution to the CDW transition stems from momentum-dependent electron-phonon coupling. We conclude that this mechanism prevents the CDW from vanishing, although the nesting conditions within the magnetically ordered phases deteriorate.
\end{abstract}
\maketitle

\section{introduction}

The interactions between the various charge and magnetic degrees of freedom are the object of an extensive investigation aimed to understand and control the related macroscopic observables \cite{Chang2012, Lu2015, Kawasaki2017, Song2021, Zeng2022, Teng2023}. 
A particularly close relationship can exist between entities that are based on the same foundation, e.g.\ the electronic structure. 
An example of the latter is charge density wave (CDW) order and magnetism driven by the Ruderman–Kittel–Kasuya–Yosida (RKKY) mechanism \cite{Ruderman1954, Kasuya1956, Yosida1957}. 
CDW, seen in quasi-low-dimensional materials, is predicted by the canonical Peierls-Fr\"{o}hlich model \cite{Peierls1955, Frohlich1954} to originate from nesting of parallel segments of the Fermi surface (FS) assisted by coupling between electrons and phonons and accompanied by crystal structure modulation. RKKY magnetism of rare earth based metals, leading even to the formation of topologically non-trivial spin textures such as magnetic skyrmions \cite{Kurumaji2019}, rely on itinerant electrons which mediate the interaction between local magnetic moments, and thus is sensitive to charge distribution in space and to the Fermi surface contour \cite{Inosov2009}. 
The relation between CDW and magnetism can show either a constructive or destructive character. 
On the one hand, the presence of a readily developed charge modulation may serve as a template for spin fluctuations and eventually govern the onset of a magnetically ordered state \cite{Salters2023}.  
On the other hand, long-range magnetic ordering, particularly ferromagnetism, competes with the charge density wave state \cite{Coelho2019, Ramakrishnan2020, Zhou2023}. 
Therefore, the number of systems in which both entities coexist is severely limited.

One of the systems in which both types of ordering occur is the $R$NiC$_2$ family, where $R$ stands for a rare earth ion. Within this group of compounds, at least two kinds of CDWs have been reported. Formation of Ni-Ni pairs entails structural modulation with modulation wave vectors : $q_1$ = (0.5, 0.5+$\eta$, 0), characteristic for early lanthanide based family members \cite{Wolfel2010, Yamamoto2013, Shimomura2016}, and commensurate $q_2$ = (0.5, 0.5, 0.5), dominating the late-lanthanide based $R$NiC$_2$ \cite{Maeda2019}. The $q_1$-type CDW has been suggested not only to coexist but also to scaffold the antiferromagnetic state (AFM), occurring in NdNiC$_2$ and GdNiC$_2$ with almost identical periodicity (propagation vector) as the CDW lattice modulation \cite{Yamamoto2013, Hanasaki2017}. The field-induced transitions to metamagnetic (MM) and field-aligned ferromagnetic (FA-FM) states yet entirely suppress this type of CDW \cite{Hanasaki2017}, which effect has also been observed at the onset of the ferromagnetic groundstate in SmNiC$_2$ \cite{Shimomura2009, Hanasaki2012}. The $q_2$-type CDW however exhibits the signatures of a distinct response to the magnetic ordering. 
Maeda \textit{et al.} \cite{Maeda2019} have demonstrated that this state remains entirely untouched upon crossing the N\'{e}el temperature and onset of the AFM phase in DyNiC$_2$, HoNiC$_2$ and ErNiC$_2$. Moreover, the arguments raised upon the transport and magnetotransport properties of polycrystalline samples of TmNiC$_2$ \cite{Kolincio2020}, HoNiC$_2$ and ErNiC$_2$ \cite{Kolincio2024} suggest that, unlike $q_1$-type CDW, the $q_2$-type survives also in the MM and FA-FM phases.  
However, the limitation of the experiments performed on a polycrystalline material is that they provide only an indirect probe of the CDW. 
While magnetoresistance and Hall responses deliver information on the carrier density and mobility,  strongly influenced by CDW via Fermi surface reconstruction, they can also be affected by different, CDW-independent factors, such as the presence of Weyl nodes, as predicted to occur in the electronic structure of $R$NiC$_2$ \cite{Ray2022}. Therefore, although the previous studies delivered substantial clues, the direct proof required to unambiguously confirm or deny the robustness of CDW order across the field induced magnetic states is missing.

Regardless of the mechanism triggering the Peierls transition to a CDW state, and its impact on the electronic structure, the essence of this state is the spatial modulation of electronic charge density, concomitant with the lattice distortion with matching periodicity \cite{Gruner1988, Monceau2012}.  The related satellite peaks, which are observable by diffraction methods are thus a sensitive probe of the CDW order parameter. They can be traced under evolving external conditions or across the boundaries of competing or intertwined states  \cite{Chang2012, Chang2016}. In this paper, we present a direct investigation of the CDW response to long-range magnetism by means of the single-crystal x-ray diffraction technique. 

\section{experimental and calculation methods}
\subsection{Crystal growth and bulk characterization}

All experiments were performed on a single crystalline sample of TmNiC$_2$ grown with the  floating zone technique, as previously described in Ref.~\cite{Roman2023}. 
Field- and temperature-dependent magnetization measurements were conducted using the Vibrating Sample Magnetometry option (VSM) of a Physical Property Measurement System (PPMS). 
Low-temperature specific heat data at zero external magnetic field and at 0.3\,T were collected with a home-made calorimeter employing an adiabatic step-heating technique.

\subsection{High temperature diffraction experiment}

To determine the temperature dependence of the amplitude of Ni atom displacements, diffraction data of a small fragment of a TmNiC$_2$ crystal were collected using Mo K$_{\alpha}$ radiation on a Bruker KAPPA APEX II diffractometer system equipped with an Oxford Cryostream cooling system. 
A dry stream of nitrogen was used in the 100-300 K range (10 K/step).
Frame data of both twin domains were reduced to intensity values using SAINT-Plus (HKLF 5 file format with reflection overlap information) and a correction for absorption effects was applied using the multi-scan approach implemented in TWINABS \cite{bruker}.
The structures were refined using SHELX using a common starting structure. 
Hundred refinement cycles resulted in full convergence in all cases.

\subsection{Synchrotron diffraction experiment}

The low temperature and high field diffraction experiment has been performed at P09 beamline of Petra III synchrotron equipped with a 14 T superconducting cryomagnet \cite{DESY} and a variable temperature insert. 
We have used a horizontal scattering geometry with a Dectris PILATUS 300k pixel detector. 
In this experimental setup, the $h=0$ main reflections diffract in the horizontal scattering plane ($\gamma=0$) \cite{DESY, DESY_S}. 
By choosing a radiation energy of 25 keV (wavelength $\lambda=0.49594$ \AA{}), the $h=\pm\frac12$ satellites diffract in a vertical out of plane angle of $\gamma=\sin^{-1}(\frac{\lambda}{2a})=4.07^\circ$ (assuming $a=3.49$ \AA{}), which is comfortably in the range of the $\pm5^\circ$ vertical opening of the x-ray window. 
The selected energy of 25 keV is away from any resonance lines related to the elements in TmNiC$_2$.

\begin{figure} [h]
  \includegraphics[angle=0,width=1.0\columnwidth]{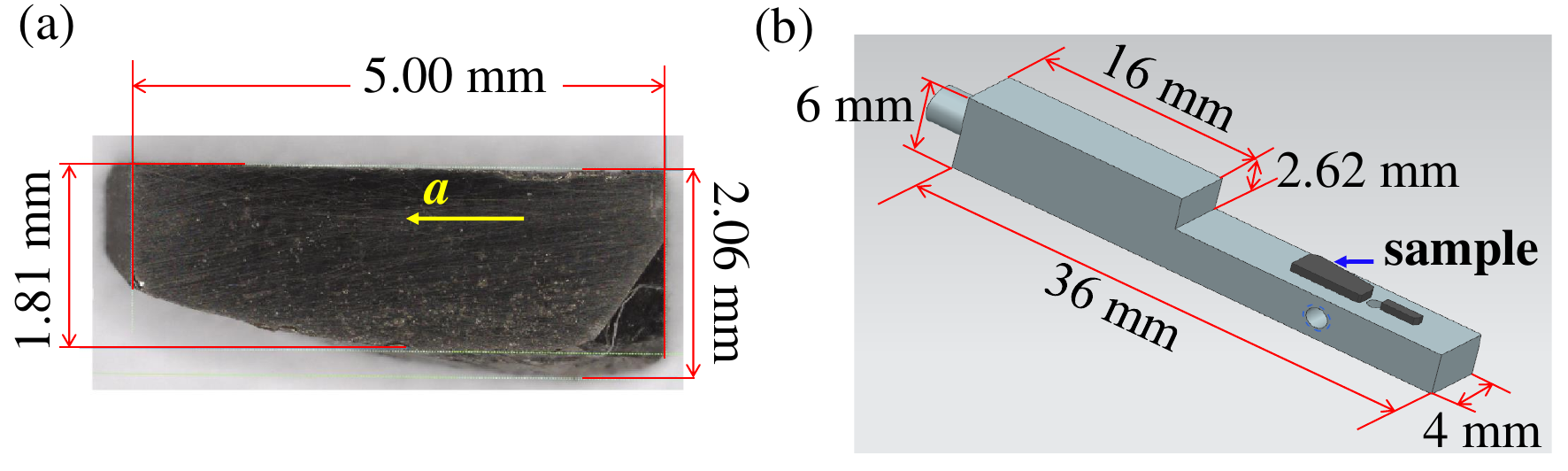}
\caption{\label{fig_sample} (a) The TmNiC$_2$ sample used for the diffraction experiment. Dimensions and crystallographic $a$-axis  direction are displayed. Sample thickness, not shown in the figure is $t$ = 0.66 mm. (b) Oxygen- free copper sample holder used for synchrotron experiment with its key dimensions. The largest surface of the sample, parallel to (011) plane, contacts the holder to maximize the heat exchange, and to ensure the mechanical stability of the sample exposed to strong magnetic field.}
  \end{figure}

Since magnetic ordering onsets at a N\'eel temperature as low as $T_{\rm N} = 5.5$ K, it is of crucial importance to ensure stable experimental conditions and to exclude the possibility that beam heating effects increase the temperature to such an extent that $T_{\rm N}$ is exceeded from below.

To facilitate heat dissipation, we have used a sample holder custom-made of highly conducting, oxygen-free copper, and mounted the sample with a thin layer of Loctite Stycast 2850 FT. The sample dimensions - area $S$ and thickness $t$, with ratio $\frac{S}{t} = $ 14.37 mm were optimized to maximize the heat contact. 
The picture of the sample and the drawing of the sample holder, which serves as a heat sink, are shown in Fig.\,\ref{fig_sample}. 
At P09 beamline, at liquid helium temperatures, and X-ray energies of $\approx$ 8 keV where the photon flux is maximum, an attenuation factor of $R_v$ 50 is usually used to ensure that the sample remains at base temperature where the sample surface is directly cooled by the pumped cold helium gas. However here, as the AFM order parameter could not be directly probed, in order to guarantee that $T_{\rm N}$ is not exceeded, 10 to 20 times higher attenuation factors have been used. For the scans at temperatures above 3 K, the beam was attenuated by a factor of $R_v$ = 500 when scanning both satellite reflections, while the Bragg peaks (0, 5, 7) and (0, 6, 8) have been measured with attenuation factor of 5000 and 1000, respectively. The scans at 2.9 K have been conducted with attenuation rate $R_v$ = 1000, and 12333 for satellite, and Bragg peaks, respectively. 
The total power carried by the beam reaching the sample is 28 $\mu$W at $R_v= 500$, and 14 $\mu$W when applying a higher attenuation factor $R_v=1000$ (see the Supplemental Material (SM)~\cite{SM} for further details). \nocite{Kriminski_2003, Mhaisekar2005, XCOM, Kim2012, Daudin_1984, cbflib} 
To estimate the maximum temperature increase $\Delta T$ at the illuminated sample spot we have considered a worst-case scenario, entirely neglecting the convection cooling mechanism and relying solely on the conduction channel. We have compared the sample geometry and thermal conductivity at relevant temperatures with the  reference experiment performed previously on TmVO$_4$ sample \cite{DESY_S}, which while illuminated by 26 $\mu$W of incident beam, was stably maintained at 1.9 K under high vacuum, and thus was cooled by conduction mechanism alone. We find that even with the most cautious approach case, $\Delta T$ is lower than 0.3 K, and sample temperature remains far below $T_{\rm N}$ (see SM \cite{SM} for more details).

\begin{figure*} [ht]
  \includegraphics[angle=0,width=2.0\columnwidth]{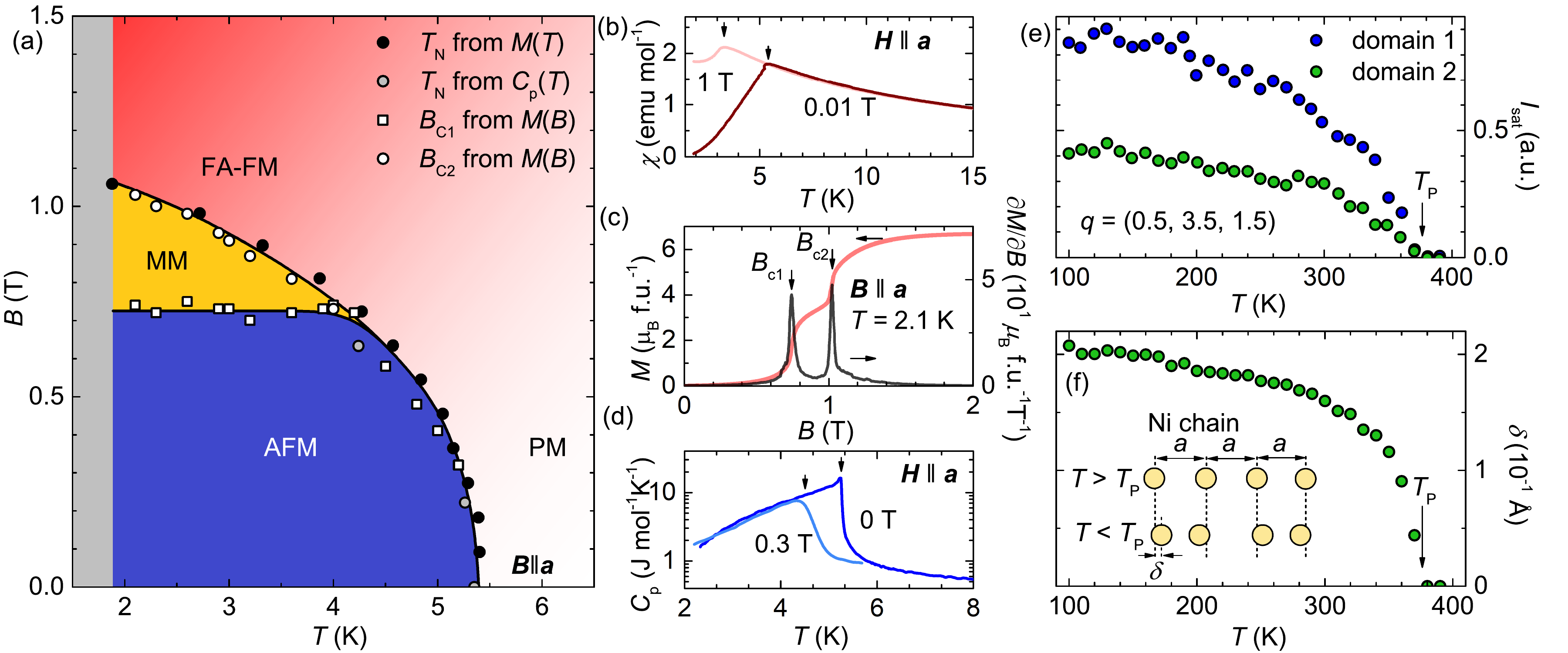}
\caption{\label{fig_diagram} Magnetism and CDW TmNiC$_2$. (a) Magnetic phase diagram including paramagnetic (PM), antiferromagnetic (AFM), field induced metamagnetic (MM), and field aligned-ferromagnetic (FA-FM) states. Phase boundaries are constructed using the transition temperatures $T_{\text{N}}$ and magnetic fields $B_{\text{c1}}$ and $B_{\text{c2}}$ obtained from temperature and field dependencies of magnetization, as depicted in panels (b) and (c) as well as temperature dependence of specific heat as shown in panel (d), respectively. Vertical arrows in panels (b) – (d) mark the transition temperatures and fields. (e) Thermal dependence of  $q$ = (0.5, 3.5, 1.5) satellite peak intensity, corresponding to two crystallographic domains. Peierls temperature $T_{\text{P}}$ as indicated by the onset of satellite intensity is marked using a vertical arrow. The presence of two twin domains, giving the same fundamental peaks, but different sets of satellite reflections is caused by twinning associated to the symmetry lowering from spacegroup \textit{Amm}2 (orthorhombic) to \textit{Cm} (monoclinic) at the CDW transition \cite{Roman2023}. (f) \textbf{\textit{a}} axis direction displacement of the Ni atoms $\delta$ from the average position in the fundamental ($T>T_{\text{P}}$) state, as sketched in the inset.}
  \end{figure*}

We emphasize that our worst-case estimate does not take into account the presence of exchange gas (He) in the sample chamber, which directly cools down the radiation exposed sample surface, and thus, more effectively minimizes the beam heating effects than heat conduction through sample and sample holder.
Moreover, our model does not capture the fraction of photons which are diffracted - not absorbed - and thus do not contribute to the heat process.
We then conclude that, the real heating effect is significantly weaker than estimated here, and the parameters chosen by us are more than sufficient to rule out any relevant beam heating effects \cite{DESY_S}.

\subsection{DFT calculations}
To compute the electronic structure and phonon dispersion we employed Density Functional Theory (DFT) and Density Functional Perturbation Theory (DFPT) calculations, as implemented in Quantum ESPRESSO \cite{Giannozzi_2009, giannozzi2017advanced}. Phonon linewidths were computed from the imaginary part of the phonon selfenergy, with Wannier-interpolated electron-phonon matrix elements, as implemented in EPW \cite{epw1, epw2, epw3}. The nesting function was computed in the constant-matrix approximation, as defined in Ref. \cite{epw2}. DFT calculations were performed using Optimized Norm-Conserving Vanderbilt pseudopotentials \cite{Hamann2013}, employed with a Perdew-Burke-Ernzerhof (PBE) exchange-correlation functional \cite{perdew1996generalized}, within the frozen-core approximation. For the self-consistent calculation of the ground-state charge density we expanded the Kohn-Sham wavefunctions in plane waves, with a cutoff of 120 Ry, and integrated over a 10$\times$10$\times$10 grid in the Brillouin zone, with a 0.005 Ry smearing on the electronic occupations. Phonon calculations were performed on a 4$\times$4$\times$4 grid of wavevectors. A satisfying wannierization of the bands close to the Fermi energy is obtained including 8 wannier functions, with an energy window up to 5.2 eV (relative to the Fermi energy), a frozen energy window from -0.2 to 1.9, and an initial guess with 5 $d$ orbitals over Ni, and 3 $sp^2$-like orbitals in the middle point between the two C atoms. 
The electron-phonon matrix elements were interpolated from the coarse 4$\times$4$\times$4 grid (for both, phonons and electrons) to a $24\times 24\times 24$ grid for electrons, and directly on the phonon dispersion path for phonons. 

All calculations on the parent phase of TmNiC$_2$ were performed using the experimental structure at 400 K, while the $q_{2c}$-type CDW phase was derived theoretically in a supercell by following the soft eigenvector in the phonon dispersion of the parent phase, which is in excellent agreement with the experimental one.

\section{results and discussion}
\subsection{CDW and magnetic ordering}

The starting point for our analysis is the careful and accurate determination of the boundaries between each magnetic phase, which leads to the construction of the $B-T$ phase diagram of TmNiC$_2$, with respect to a field applied along the  crystallographic \textit{\textbf{a}} axis, which is the easy direction of magnetization.    
As depicted in Fig.~\ref{fig_diagram}(a), the magnetic dome of TmNiC$_2$ contains three different types of magnetic orders, which are consistent with the preliminary diagram proposed by Kohsikawa \textit{et al.}~\cite{Koshikawa1997}, which is complemented and further developed here. 
Apart from the high temperature paramagnetic (PM) state (white color background), one can distinguish the antiferromagnetic ground state (blue) as well as metamagnetic (yellow) and field-aligned ferromagnetic (red) states. 
The temperatures and magnetic fields corresponding to the transitions between the subsequent phases are obtained by means of bulk magnetic and thermal properties, respectively. 
The temperature dependencies of the magnetic susceptibility $\chi(T)$ and specific heat $C_{\text{p}}(T)$ show pronounced maxima at the N\'{e}el temperature $T_{\text{N}}$ as depicted in Fig.~\ref{fig_diagram}(b) and (d) while the field dependence of magnetization $M(B)$ (Fig.~\ref{fig_diagram}(c)) reveals the field induced transitions to MM and FA-FM states manifesting as inflections at $B_{\text{c1}}$ and $B_{\text{c2}}$, respectively. The precise position of these critical fields is determined from the maxima of the magnetization derivative with respect to the field, $\partial M / \partial B$. The field dependencies of all quantities have been corrected for the demagnetization factor: $B=\mu_0H-NM$, where $\mu_0H$ is the applied field and $N$ relates to the sample shape \cite{Aharoni_1998}. 
Further details are included in the SM \cite{SM}.

Figures~\ref{fig_diagram}(e) and \ref{fig_diagram}(f) demonstrate the onset and development of $q_2$-type CDW, based on diffraction data obtained using laboratory diffractometer Bruker KAPPA APEX II. 
For further details on the experimental procedure and data refinement see Ref.~\cite{Roman2023} and SM \cite{SM}. 
The thermal dependence of the $q = (0.5, 3.5, 1.5)$ satellite reflection intensity is shown in panel (e), while panel (f) exhibits the temperature dependence of Ni atoms displacement $\delta$ from their positions along \textit{\textbf{a}} - axis in the fundamental state (at $T>T_{\text{P}}$). 
Both $I_{sat}(T)$ and $\delta(T)$ saturate at lowest temperatures. 
Remarkably, the latter quantity, derived from single crystal structure analysis, reaches a relatively large value of $\approx 0.21$ \AA, corresponding to $\approx 6$ \% of the average Ni-Ni distance. 
This observation is a characteristic landmark for the unconventional mechanisms  \cite{McMillan1975, Friend_1979} participating in CDW formation. 
In addition to the mechanism of Fermi surface nesting, the stronger electron-phonon coupling \cite{Varma_1983, Galli2002} seems to be involved in the stabilization of this state. 

\subsection{Character of CDW transition}

\begin{figure} [h!]
  \includegraphics[angle=0,width=0.9\columnwidth]{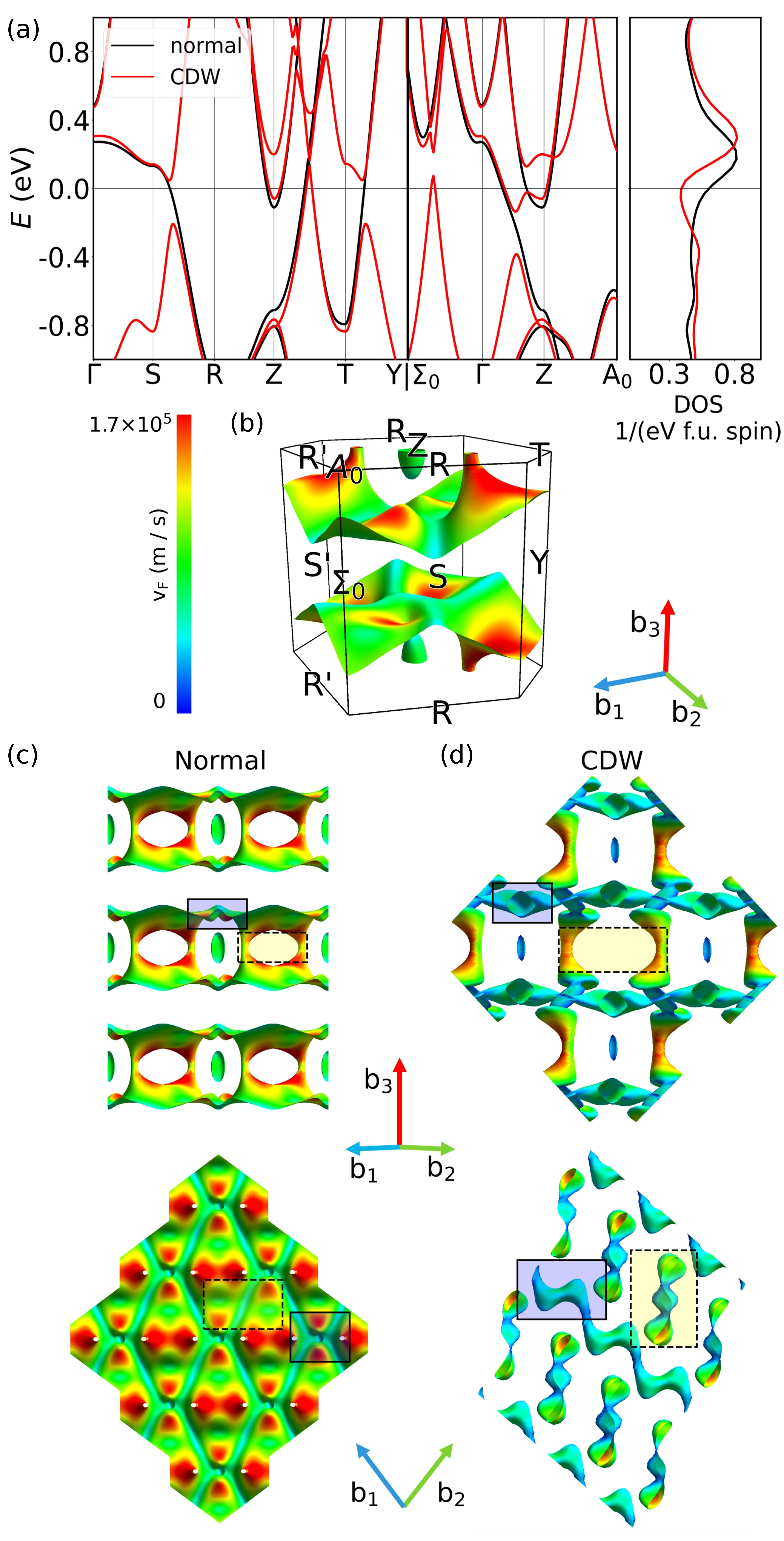}
\caption{\label{bands+FS}(a) Electronic band structure and DOS of $\mathrm{TmNiC_{2}}$ for both the normal phase (black line) and the $q_{2c}$-type CDW phase (red line). The Fermi energy is taken as zero. Figure (b) shows the Fermi surface in the first Brillouin zone (FBZ) for the normal phase, along with the special points.  Panel (c) shows the Fermi surface for the normal phase from different perspectives (above, a lateral view; below, a top view). Similarly, (d) shows the Fermi surface for the $q_{2c}$-type CDW phase. The solid and dashed black boxes indicate corresponding features in the Fermi surface of the two phases. The color scale represents the Fermi velocity. The high-symmetry \textbf{k} points in the FBZ are $\Gamma=(0,0,0)$, $\text{S}=(0,0.5,0)$, $\text{R}=(0,0.5,0.5)$, $\text{Z}=(0,0,0.5)$, $\text{T}=(-0.5,0.5,0.5)$, $\text{Y}=(-0.5,0.5,0)$, $\Sigma_{0}=(0.39,0.39,0)$ and $\text{A}_{0}=(0.39,0.39,0.5)$.}
\end{figure}

The strength of the electron-phonon mechanism relative to the Fermi surface nesting can be directly compared by means of Density Functional Theory calculations \cite{Johannes2008, lian2019coexistence}. In Fig. \ref{bands+FS}(a) we show the band structure of the parent and $q_{2c}$-type CDW phases following the same path in reciprocal space. In the parent phase, the bands cross the Fermi energy in the S--R, Z--T, T--Y, and $\Gamma$--Z lines 
(for the full band structure in a wider energy range we refer the reader to Fig.~S7 of SM \cite{SM}).  

As shown in Fig.~\ref{bands+FS}(a), the bands within the CDW phase exhibit the formation of 250 meV splittings near the Fermi level in the S--R, T--Y, and $\Gamma$--Z lines. Together these lead to a pseudogap in the density of states (DOS). Going from the parent to the CDW phase, the DOS decreases from 0.605 to 0.408 states/(eV f.u. spin). The energy window in which the DOS of the CDW phase is lower than the parent phase, which we define as the CDW gap, is estimated to be 495 meV wide (see Fig. S8 of SM \cite{SM}).

In Fig.~\ref{bands+FS}(c) and \ref{bands+FS}(d) we show the Fermi surface (FS) of the parent (normal) state, and within the $q_{2c}$-type CDW state, respectively. Both the unperturbed and distorted state FS show large similarities to those reported for YNiC$_2$ \cite{Roman_2024_Y}, and LuNiC$_2$ \cite{Steiner2018}. 
The main parts of the normal state FS are quasi-planar sheets running perpendicularly to the \textbf{b}$_3$ direction, with X-shaped valleys, as seen in the top view of Fig. \ref{bands+FS}(c).
The Fermi surface also contains other characteristic features: hourglass-shaped columns connecting the sheets (highlighted with dashed black line rectangles in the top panel of Fig. \ref{bands+FS}(c)), and egg-shaped pockets. 
While CDW transition removes the majority of the quasi-planar FS component, as demonstrated in Fig.~\ref{bands+FS}(d), leaving only a sine-like sheet (highlighted with a solid black line bordered rectangle in the lower panel of Fig. \ref{bands+FS}(d)), the two smaller elements - columns and egg-shaped pockets - remain untouched. 
The Fermi surface decomposition into a set of partially connected or
isolated pockets with dispersion in all three dimensions is consistent with the reported previously \cite{Roman2023} loss of transport properties anisotropy upon lowering the temperature below $T_P$ = 319 K.

\begin{figure} [h]
  \includegraphics[angle=0,width=1.0\columnwidth]{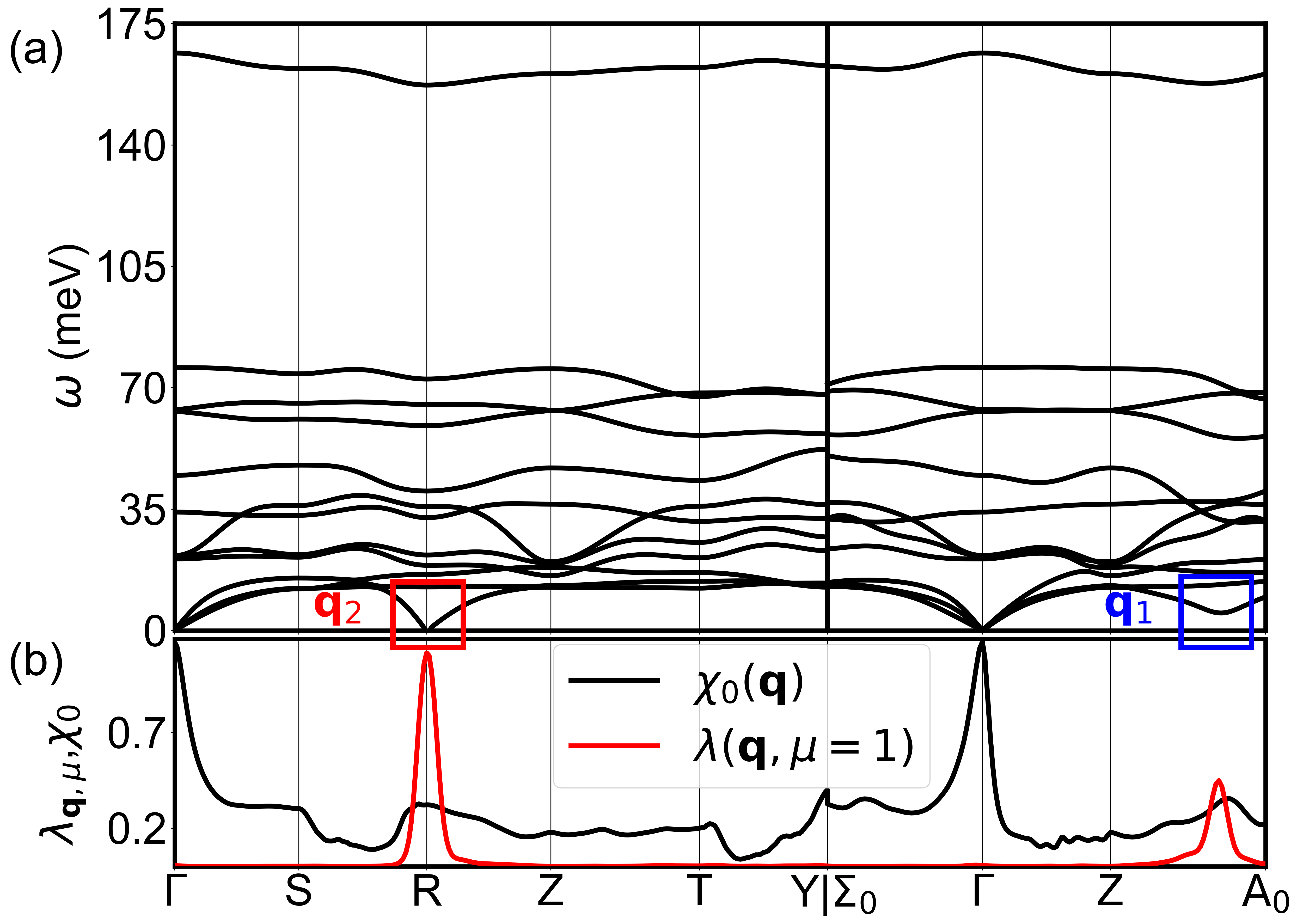}
\caption{\label{PH_disp+lw} (a) Phonon dispersion of $\mathrm{TmNiC_{2}}$ revealing softening at the wave vectors $q_{2}$ and $q_{1}$, highlighted as a red and blue box, respectively. (b) Electron-phonon coupling  $\lambda_{\textbf{q},\mu}$ (red line) and nesting function $\chi_{0}$ (black line).}
  \end{figure}

To complement the electronic structure and its evolution, we proceed to discuss the vibrational properties and their interactions with the crystal lattice. In Fig. \ref{PH_disp+lw}(a) we show the phonon dispersion for the normal phase of TmNiC$_2$; The phonon dispersion branch exhibits two dips corresponding to the wave vectors R = $q_{2} = (0.5, 0.5, 0.5)$, and $q_{1} = (0.5, 0.56, 0)$, which is located between $Z$ and $A_{0}$. 
The softening at $q_2$ is visibly stronger than the one at $q_1$, and the mode reaches an imaginary frequency at this point, being a hallmark of a phonon instability. 
The significantly weaker dip at $q_1$ suggests that the potential instability at this wavevector remains latent. 

Armed with the details of electronic and phonon structures, we can directly examine the relative role of nesting and electron-phonon coupling strength, by comparing the relevant quantities: nesting function $\chi_{0}(\textbf{q})$, defined in the static limit as: 
\begin{equation}
    \chi_{0}(\textbf{q}) = \sum_{\textbf{k}} \delta(\epsilon_{\textbf{k}}-\epsilon_{F})\delta(\epsilon_{\textbf{k}+\textbf{q}}-\epsilon_{F}),
\end{equation} 

where $\epsilon_{F}$ is the Fermi energy while $\epsilon_{\textbf{k}}$ and $\epsilon_{\textbf{k}+\textbf{q}}$ are the energies of the electronic states at the wavevectors $\textbf{k}$ and $\textbf{k} +\textbf{q}$; and anisotropic electron-phonon coupling $\lambda_{\textbf{q},\mu}$, with $\mu$ being the phonon mode. 
The nesting function and anisotropic coupling for the softest mode ($\mu = 1$) are shown in Fig.~\ref{PH_disp+lw}(b). 
For the two wave vectors corresponding to the soft modes ($q_1$ and $q_2$), we find peaks in both, $\chi_{0}(\textbf{q})$ and $\lambda_{\textbf{q},\mu}$. 
However, the broadened and weak maxima of $\chi_{0}(\textbf{q})$, similar to those of LaNiC$_2$, which shows no CDW, contrast with the strong and well-defined peaks seen in the cases of NdNiC$_2$, SmNiC$_2$, and GdNiC$_2$ \cite{Laverock2009}, indicate that nesting alone is not sufficient to drive the CDW transition in TmNiC$_2$.
Moreover, the nesting function is also as large for other wavevectors (e.g. Y and $\Sigma_{0}$), for which we find no softening. 
In contrast to the weak and broadened peaks in $\chi_{0}(\textbf{q})$, the electron-phonon coupling shows a sharp maximum at $q_2$, which corresponds to the observed CDW-related satellite peaks, and a smaller one at $q_1$. 
This result suggests that the electron-phonon coupling is the main mechanism for the CDW transition, which is consistent with strong the lattice distortion and the high intensity of satellite peaks.
In other words, Fermi surface nesting has little predictive power and plays a lesser role compared to electron-phonon coupling.

\subsection{Interplay between CDW and magnetism}

Knowing that the momentum dependent electron-phonon coupling plays a decisive role in CDW formation, we may now examine, whether this mechanism can defend the CDW from being suppressed by long range magnetism. 
To explore the interaction between charge density waves and phases present in the magnetic phase diagram of TmNiC$_2$ (see Fig.~\ref{fig_diagram}(a)), we performed synchrotron diffraction experiments at low temperatures ($T < 12$ K) and strong magnetic fields ($B\leq 3$\,T), which allowed us to tune the magnetic state.

\begin{figure} [t!]
  \includegraphics[angle=0,width=1.0\columnwidth]{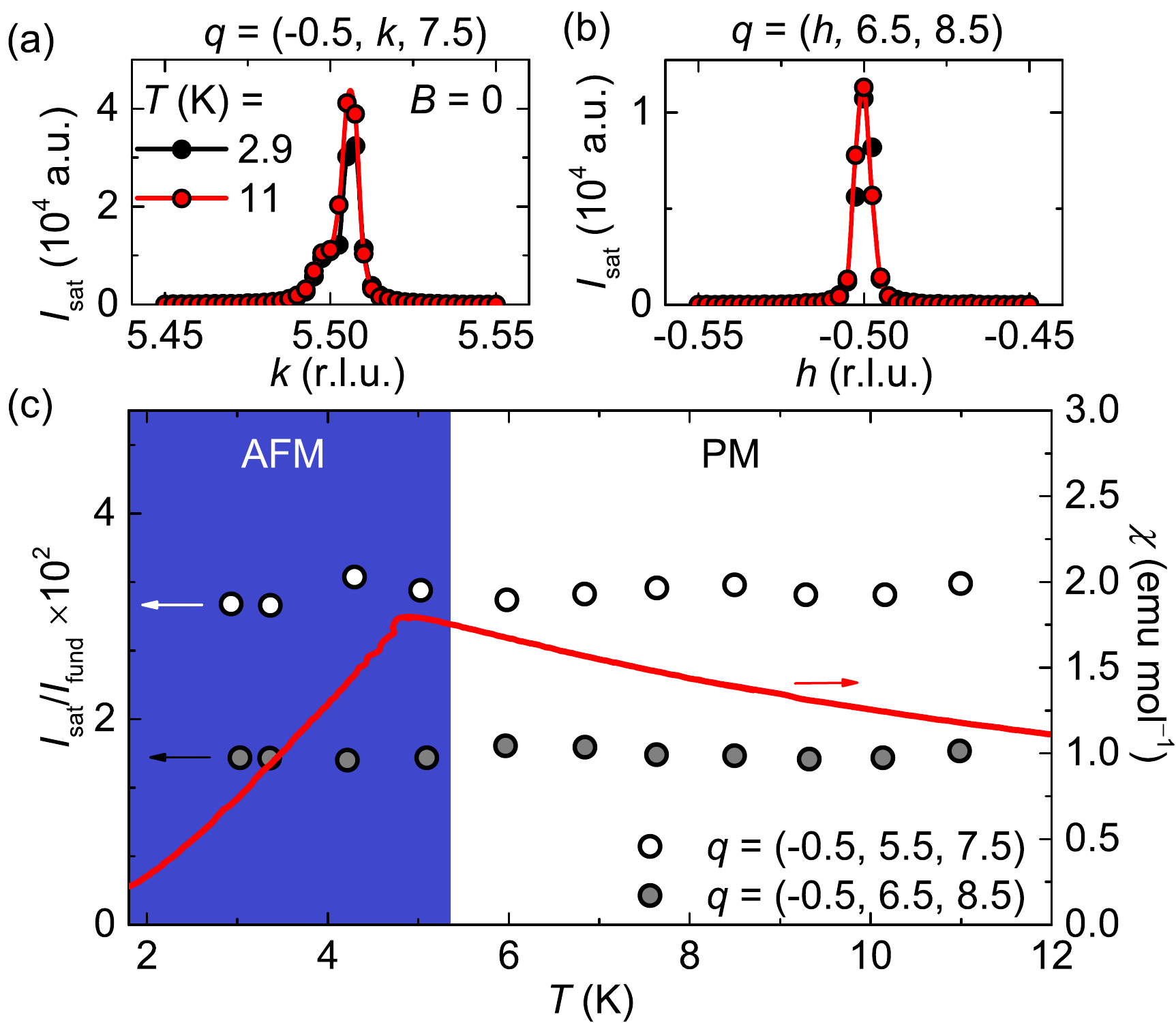}
\caption{\label{fig_temperature} CDW satellites across the antiferromagnetic transition. (a), (b): Reciprocal space profiles of (-0.5, 5.5, 7.5) [panel (a)] and (-0.5, 6.5, 8.5) [panel (b)] peaks at temperatures above and below $T_{\text{N}}$: at $T=11$ K (red color) and $T=2.9$ K (black), respectively. (c): thermal dependence of these peaks intensity normalized by the neighboring fundamental peaks (0, 5, 7)  and (0, 6, 8), respectively (white and gray points respectively, left hand scale) at the vicinity of $T_{\text{N}}$ at zero applied field. Temperature dependence of magnetic susceptibility $\chi(T)$ measured at $\mu_0H = 0.01$ T (red line, right hand scale) illustrates the transition from PM (white background) to AFM (blue background) state.}
  \end{figure}

During the low-temperature diffraction experiment, we have selected and followed two satellite reflections $q = (-0.5, 5.5, 7.5)$ and $q = (-0.5, 6.5, 8.5)$.
Their nearest-neighbor fundamental Bragg peaks (0, 5, 7) and (0, 6, 8) have been used as reference. 
Figures~\ref{fig_temperature}(a) and \ref{fig_temperature}(b) depict the representative peak profiles, along $k$ and $h$ reciprocal lattice directions, 
measured at stable temperatures of $T = 11$ K (red color) and  2.9 K (black color), respectively. 
Within the experimental resolution, the reflections show the same shape intensity at both temperatures, standing above and below $T_{\text{N}}$, thus situated in PM and AFM phases, respectively. The zero field thermal dependence of satellite peak intensity normalized by the closest fundamental reflection is presented in Fig. \ref{fig_temperature}(c), together with magnetic susceptibility, in which pronounced maximum marks the magnetic transition. The courses of both $I_{\text{sat}}/I_{\text{fund}}(T)$  show similar, flat shape, with no visible drop below $T_{\text{N}}$. 
This result stands in contrast with the behavior of the $q_1$-type CDW which in NdNiC$_2$ and GdNiC$_2$ shows $\approx 50$\% and $\approx 20$\%  intensity drop at the onset of the AFM state, respectively \cite{Shimomura2016, Yamamoto2013}. 
However, our observation reproduces the behavior reported for other $R$NiC$_2$ exhibiting CDW with the same $q_2$-type of modulation - namely DyNiC$_2$, HoNiC$_2$ and ErNiC$_2$ \cite{Maeda2019}, suggesting that robustness to the antiferromagnetic ground state is a common feature of these compounds.

\begin{figure*} [t!]
  \includegraphics[angle=0,width=2.0\columnwidth]{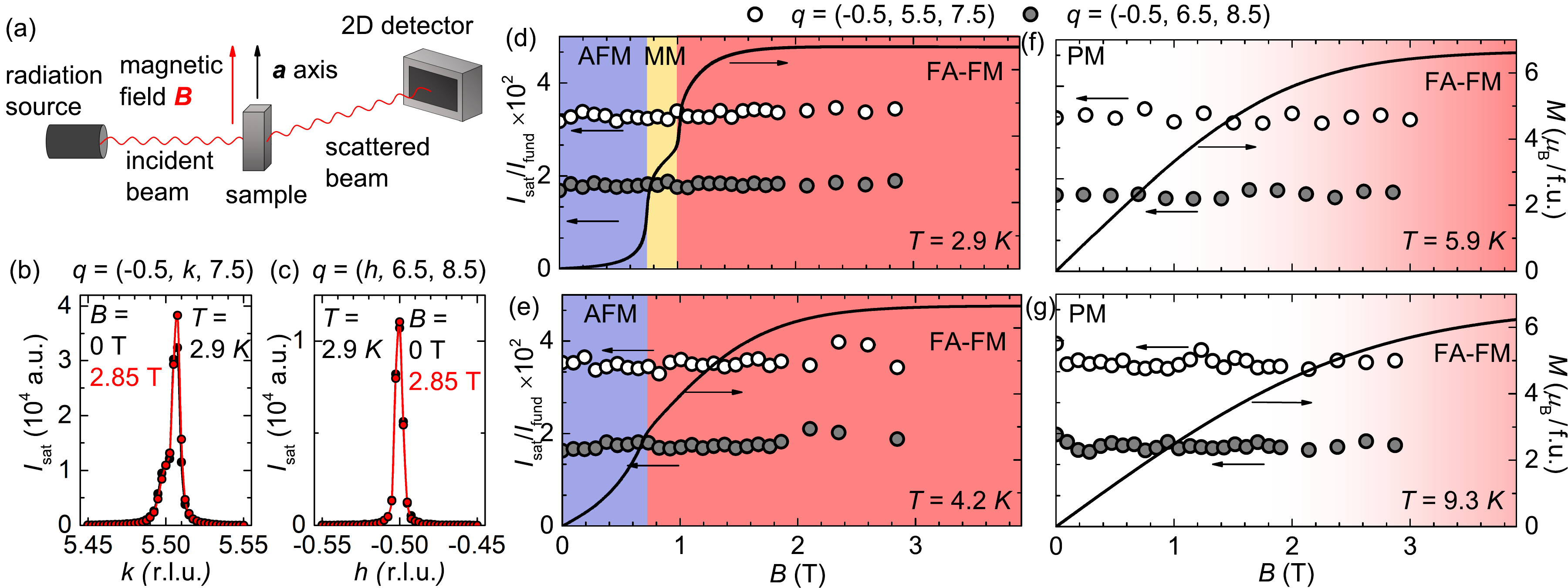}
\caption{\label{fig_field} CDW satellite peaks across the field-induced transitions. (a) Sketch of the synchrotron diffraction experiment geometry. Regardless of the scattering angle or detector position, the magnetic field (red arrow) is maintained along the \textbf{\textit{a}} crystallographic axis (black arrow). (b), (c): Reciprocal space profiles of (-0.5, 5.5, 7.5) [panel (b)], and  (-0.5, 6.5, 8.5) [panel (c)] peaks at zero field (black color) and $B=2.85$ T (red color), respectively.
Panels (d)-(g) show the field dependence of the intensity (white and gray points respectively, left hand scale) of the same peaks as in (b) and (c), normalized by the neighboring fundamental reflections [(0, 5, 7) and (0, 6, 8), respectively]. For clarity, field dependence of magnetization is also plotted with a black line (right hand scale). Labels and background colors indicate the relevant magnetic states: PM (white), AFM (blue), MM (yellow) and FA-FM (red) (see Fig. 1 for more details).}
  \end{figure*}
  
Confirming the coexistence of the $q_2$-type CDW with the antiferromagnetic groundstate in TmNiC$_2$ opens a possibility and motivation to verify the hypothesis that CDW coexists also with the remaining ingredients of the magnetic phase diagram. 
With this aim, we have followed the same satellite reflections as a function of magnetic field applied along the \textit{\textbf{a}} crystallographic axis (see a sketch in Fig.~\ref{fig_field}(a)), to subsequently scan from PM or AFM through the MM to FA-FM states. 
Figures~\ref{fig_field}(b) and \ref{fig_field}(c) present the comparison of  the reciprocal lattice profiles of the two explored satellite peaks, both measured at the lowest reached temperature of $T  = 2.9$ K: at zero field condition (black color), corresponding to the AFM groundstate, and $B= 2.85$ T (red), situated deeply inside the FA-FM state region. 
Within the measurement resolution, no significant differences can be observed between the results obtained under these two conditions: the experiment neither reveals any substantial broadening, nor weakening of the satellite peaks intensity, nor peak shifting. Furthermore, to exclude the structural changes at high fields and low temperatures, two $20^\circ\omega$ scans
(rotation of the crystal about the $\mathbf a$ axis) with two $2\theta$ positions of the detector
were performed at 2 T and 2 K. Owing to the 5$^\circ$ x-ray window of the magnet
(corresponding to more than a $h=-\frac12\to\frac12$ range) and the use of a 2D detector, multiple Brillouin zones were
thus recorded. Only the expected $h=0$ main reflections and satellite $h=\pm\frac12$ reflections were observed.

For a broader picture, in Fig. \ref{fig_field} (c)-(g) we have plotted the field dependence of the intensity of both satellites, normalized  by the neighboring fundamental peaks $I_{\text{sat}}/I_{\text{fund}}(B)$, measured at various temperatures. The displayed course of magnetization $M(B)$ measured at the same temperature, as well as the background color, corresponding to the relevant regions of the diagram from Fig. \ref{fig_diagram}(a) serve as a guide to assign the $B-T$ conditions to the underlying magnetic phases.  Within the scan resolution, $I_{\text{sat}}/I_{\text{fund}}(B)$ for both measured satellites shows the same flat characteristics in the entire experimental range, with no drop or even a less pronounced downturn, regardless of the position in the magnetic phase diagram. Conservation of both the peak shape and its initial intensity indicates that, both amplitude and spatial coherence length of the charge density wave respectively, are preserved.

\subsection{Discussion}
This observation stands in even greater contrast to the response shown by the $q_1$-type CDW.
Under similar circumstances - upon a transition to FM groundstate, as well as field induced MM or FA-FM transitions, $q_1$-type CDW suffers a rapid and complete obliteration \cite{Shimomura2009, Hanasaki2012, Hanasaki2017}.
Such a striking dissimilarity of the responses shown by these two CDW species suggests that the reason standing behind this observation is rooted at their origin. The conduction bands splitting, reaching the values at the order of charge density wave gap \cite{Kim_2013} and consequent deterioration of Fermi nesting conditions has been suggested to be the main mechanism causing the abrupt suppression of $q_1$-type CDW at the onset of the FM groundstate in SmNiC$_2$ \cite{Shimomura2009}. 
The theoretical calculations performed for LuNiC$_2$ predict the $q_2$-CDW to open a partial gap at the order of $2\Delta_{\rm CDW} \approx 300$ meV \cite{Steiner2018}. 
The same work estimates the maximum bands splitting due to the asymmetric spin-orbit coupling in this non-magnetic sibling of TmNiC$_2$ at $\approx 74$ meV, and a similar value of $\approx 84$ meV has been calculated by Hase \textit{et al.} \cite{Hase2009} in YNiC$_2$. 
Despite these models do not capture the ferromagnetic bands splitting amplification as expected in the FA-FM state of TmNiC$_2$, their results can serve as a useful guide. 
Even if greatly underestimated, the calculated magnitude of the band splitting remains a sizable fraction of the CDW band gap. Note that the saturation magnetization in FA-FM state (see Figs.~\ref{fig_diagram}(c) and \ref{fig_field}) is in TmNiC$_2$ as high as in GdNiC$_2$ ($\approx 6 \mu_B$/f.u.~\cite{Hanasaki_2011}) and an order of magnitude higher than in SmNiC$_2$ ($<$ 0.5 $\mu_B$/f.u. \cite{Onodera1998}). Therefore one expects in TmNiC$_2$ at least the same or significantly stronger band splitting than in these compounds in which $q_1$-CDW is suppressed in FA-FM and FM states, respectively. This comparison suggests that, similarly to the case of SmNiC$_2$, the Fermi surface geometry is in TmNiC$_2$ anticipated to be largely affected, yet unexpectedly, $q_2$-type CDW does not experience any, even weak, destructive consequences.

In order to become immune to the deterioration of nesting conditions and resulting CDW suppression, the $q_2$-type CDW must be stabilized by an additional factor. 
A mechanism that complements conventional Fermi surface nesting is associated with lattice degrees of freedom and is realized by the $q$-dependent electron-phonon coupling, which, if strong enough, can generate periodic lattice distortions. 
This is a way to stabilize the charge density wave \cite{Eiter_2013, Zhu2015, Zhu_2017} with a less decisive contribution from the electronic structure \cite{Johannes2008}. 
The important lattice component in TmNiC$_2$ is indicated by the pronounced transformation of the crystal structure at $T_{\text{P}}$, accompanied with symmetry descent and a significant anomaly of the specific heat \cite{Roman2023}. 
This scenario is further supported by the large intensity of the satellite peaks and strong lattice distortion revealed by our high-temperature diffraction experiment. 

Further arguments arise from our calculations of the electronic and phononic structures, which point to a major contribution of momentum-dependent electron-phonon coupling to the charge density wave transition. 
This result is in agreement with reports devoted to the characterization and the origin of CDW states in YNiC$_2$ and LuNiC$_2$ - non-magnetic siblings of TmNiC$_2$. 
The conclusions raised by Steiner \textit{et al.}\,\cite{Steiner2018}, based on the calculated electronic structure of LuNiC$_2$, cast a shadow of doubt on nesting being solely responsible for the Peierls transition to $q_2$-type CDW state. 
The analogous scenario is realized in YNiC$_2$, where the exploration of the electronic and phononic structures revealed that the decisive role in the stabilization of this state is played by the lattice degrees of freedom \cite{Roman_2024_Y}. 
Thus, as anticipated, TmNiC$_2$, which in the $R$NiC$_2$ phase diagram is situated between heavier LuNiC$_2$, and lighter YNiC$_2$ \cite{Roman_2024_Y, Steiner2018}, shares the main features of the electronic and phononic structures as well as the common roots of the $q_2$-type CDW state.

\section{conclusions}
In summary, we have performed a low temperature and high magnetic field diffraction experiment to examine the charge density wave response through all the magnetic states comprised in the phase diagram of TmNiC$_2$. 
By following the CDW-related satellite reflections on the path from paramagnetic, through antiferromagnetic and metamagnetic to field-aligned ferromagnetic states, and observing no signatures of even a weak erosion of their intensity, shape, or position, we directly demonstrate the evidence for the coexistence of $q_2$-type CDW and all these types of long range magnetism. 
We additionally confirm the robustness of a CDW-modulated state by reciprocal lattice scans ruling out the appearance of additional superstructure reflections in any of the relevant magnetic phases. 
In striking contrast to the behavior of $q_1$-CDW in early lanthanide $R$NiC$_2$, the $q_2$-type CDW in TmNiC$_2$ remains unaffected even in the presence of aligned Ising-type magnetic moments with bulk magnetization as large as 7 $\mu_B$/f.u. By DFT calculations, we also confirm the major contribution from momentum dependent electron-phonon coupling into the CDW transition. 
A plausible explanation of the vastly divergent response to magnetic ordering by these two CDW types in the $R$NiC$_2$ family - the former being sensitive and the latter immune - to electronic band splitting, is their distinct foundation, with dominant inputs from Fermi surface nesting and coupling with the crystal lattice, respectively.
\\
\begin{acknowledgments}
The project is co-financed by a bilateral project of the Polish National Agency for Academic Exchange (NAWA) Grant PPN/BAT/2021/1/00016 and Austrian Academic Exchange Service \"{O}AD Grant PL04/2022. 
M.R. acknowledges the financial support by NAWA grant BPN/BEK/2021/1/00245/DEC/1. S.\,D.\,C.\ acknowledges funding from the European Union - NextGenerationEU under the Italian Ministry of University and Research (MUR), “Network 4 Energy Sustainable Transition - NEST” project (MIUR project code PE000021, Concession Degree No.~1561 of Oct.\,11, 2022)  CUP C93C22005230007. 
We acknowledge DESY (Hamburg, Germany), a member of the Helmholtz Association HGF, for the provision of experimental facilities. Parts of this research were carried out at P09 beamline of PETRA III synchrotron.
Beamtime was allocated for proposal I-20230121 EC. 
The computational results presented have been achieved using the Vienna Scientific Cluster (VSC).
 \end{acknowledgments}
 %\bibliography{HE_bibliography}

\begin{thebibliography}{66}%
\makeatletter
\providecommand \@ifxundefined [1]{%
 \@ifx{#1\undefined}
}%
\providecommand \@ifnum [1]{%
 \ifnum #1\expandafter \@firstoftwo
 \else \expandafter \@secondoftwo
 \fi
}%
\providecommand \@ifx [1]{%
 \ifx #1\expandafter \@firstoftwo
 \else \expandafter \@secondoftwo
 \fi
}%
\providecommand \natexlab [1]{#1}%
\providecommand \enquote  [1]{``#1''}%
\providecommand \bibnamefont  [1]{#1}%
\providecommand \bibfnamefont [1]{#1}%
\providecommand \citenamefont [1]{#1}%
\providecommand \href@noop [0]{\@secondoftwo}%
\providecommand \href [0]{\begingroup \@sanitize@url \@href}%
\providecommand \@href[1]{\@@startlink{#1}\@@href}%
\providecommand \@@href[1]{\endgroup#1\@@endlink}%
\providecommand \@sanitize@url [0]{\catcode `\\12\catcode `\$12\catcode
  `\&12\catcode `\#12\catcode `\^12\catcode `\_12\catcode `\%12\relax}%
\providecommand \@@startlink[1]{}%
\providecommand \@@endlink[0]{}%
\providecommand \url  [0]{\begingroup\@sanitize@url \@url }%
\providecommand \@url [1]{\endgroup\@href {#1}{\urlprefix }}%
\providecommand \urlprefix  [0]{URL }%
\providecommand \Eprint [0]{\href }%
\providecommand \doibase [0]{https://doi.org/}%
\providecommand \selectlanguage [0]{\@gobble}%
\providecommand \bibinfo  [0]{\@secondoftwo}%
\providecommand \bibfield  [0]{\@secondoftwo}%
\providecommand \translation [1]{[#1]}%
\providecommand \BibitemOpen [0]{}%
\providecommand \bibitemStop [0]{}%
\providecommand \bibitemNoStop [0]{.\EOS\space}%
\providecommand \EOS [0]{\spacefactor3000\relax}%
\providecommand \BibitemShut  [1]{\csname bibitem#1\endcsname}%
\let\auto@bib@innerbib\@empty
%</preamble>
\bibitem [{\citenamefont {Chang}\ \emph {et~al.}(2012)\citenamefont {Chang},
  \citenamefont {Blackburn}, \citenamefont {Holmes}, \citenamefont
  {Christensen}, \citenamefont {Larsen}, \citenamefont {Mesot}, \citenamefont
  {Liang}, \citenamefont {Bonn}, \citenamefont {Hardy}, \citenamefont
  {Watenphul}, \citenamefont {Zimmermann}, \citenamefont {Forgan},\ and\
  \citenamefont {Hayden}}]{Chang2012}%
  \BibitemOpen
  \bibfield  {author} {\bibinfo {author} {\bibfnamefont {J.}~\bibnamefont
  {Chang}}, \bibinfo {author} {\bibfnamefont {E.}~\bibnamefont {Blackburn}},
  \bibinfo {author} {\bibfnamefont {A.~T.}\ \bibnamefont {Holmes}}, \bibinfo
  {author} {\bibfnamefont {N.~B.}\ \bibnamefont {Christensen}}, \bibinfo
  {author} {\bibfnamefont {J.}~\bibnamefont {Larsen}}, \bibinfo {author}
  {\bibfnamefont {J.}~\bibnamefont {Mesot}}, \bibinfo {author} {\bibfnamefont
  {R.}~\bibnamefont {Liang}}, \bibinfo {author} {\bibfnamefont {D.~A.}\
  \bibnamefont {Bonn}}, \bibinfo {author} {\bibfnamefont {W.~N.}\ \bibnamefont
  {Hardy}}, \bibinfo {author} {\bibfnamefont {A.}~\bibnamefont {Watenphul}},
  \bibinfo {author} {\bibfnamefont {M.~v.}\ \bibnamefont {Zimmermann}},
  \bibinfo {author} {\bibfnamefont {E.~M.}\ \bibnamefont {Forgan}},\ and\
  \bibinfo {author} {\bibfnamefont {S.~M.}\ \bibnamefont {Hayden}},\ }\bibfield
   {title} {\bibinfo {title} {Direct observation of competition between
  superconductivity and charge density wave order in
  {YBa$_2$Cu$_3$O$_{6.67}$}},\ }\href {https://doi.org/10.1038/nphys2456}
  {\bibfield  {journal} {\bibinfo  {journal} {Nature Physics}\ }\textbf
  {\bibinfo {volume} {8}},\ \bibinfo {pages} {871} (\bibinfo {year}
  {2012})}\BibitemShut {NoStop}%
\bibitem [{\citenamefont {Lu}\ \emph {et~al.}(2015)\citenamefont {Lu},
  \citenamefont {Wang}, \citenamefont {Wu}, \citenamefont {Wu}, \citenamefont
  {Zhao}, \citenamefont {Zeng}, \citenamefont {Luo}, \citenamefont {Wu},
  \citenamefont {Bao}, \citenamefont {Zhang}, \citenamefont {Huang},
  \citenamefont {Huang},\ and\ \citenamefont {Chen}}]{Lu2015}%
  \BibitemOpen
  \bibfield  {author} {\bibinfo {author} {\bibfnamefont {X.~F.}\ \bibnamefont
  {Lu}}, \bibinfo {author} {\bibfnamefont {N.~Z.}\ \bibnamefont {Wang}},
  \bibinfo {author} {\bibfnamefont {H.}~\bibnamefont {Wu}}, \bibinfo {author}
  {\bibfnamefont {Y.~P.}\ \bibnamefont {Wu}}, \bibinfo {author} {\bibfnamefont
  {D.}~\bibnamefont {Zhao}}, \bibinfo {author} {\bibfnamefont {X.~Z.}\
  \bibnamefont {Zeng}}, \bibinfo {author} {\bibfnamefont {X.~G.}\ \bibnamefont
  {Luo}}, \bibinfo {author} {\bibfnamefont {T.}~\bibnamefont {Wu}}, \bibinfo
  {author} {\bibfnamefont {W.}~\bibnamefont {Bao}}, \bibinfo {author}
  {\bibfnamefont {G.~H.}\ \bibnamefont {Zhang}}, \bibinfo {author}
  {\bibfnamefont {F.~Q.}\ \bibnamefont {Huang}}, \bibinfo {author}
  {\bibfnamefont {Q.~Z.}\ \bibnamefont {Huang}},\ and\ \bibinfo {author}
  {\bibfnamefont {X.~H.}\ \bibnamefont {Chen}},\ }\bibfield  {title} {\bibinfo
  {title} {Coexistence of superconductivity and antiferromagnetism in
  {(Li$_{0.8}$Fe$_{0.2}$)OHFeSe}},\ }\href {https://doi.org/10.1038/nmat4155}
  {\bibfield  {journal} {\bibinfo  {journal} {Nature Materials}\ }\textbf
  {\bibinfo {volume} {14}},\ \bibinfo {pages} {325} (\bibinfo {year}
  {2015})}\BibitemShut {NoStop}%
\bibitem [{\citenamefont {Kawasaki}\ \emph {et~al.}(2017)\citenamefont
  {Kawasaki}, \citenamefont {Li}, \citenamefont {Kitahashi}, \citenamefont
  {Lin}, \citenamefont {Kuhns}, \citenamefont {Reyes},\ and\ \citenamefont
  {Zheng}}]{Kawasaki2017}%
  \BibitemOpen
  \bibfield  {author} {\bibinfo {author} {\bibfnamefont {S.}~\bibnamefont
  {Kawasaki}}, \bibinfo {author} {\bibfnamefont {Z.}~\bibnamefont {Li}},
  \bibinfo {author} {\bibfnamefont {M.}~\bibnamefont {Kitahashi}}, \bibinfo
  {author} {\bibfnamefont {C.~T.}\ \bibnamefont {Lin}}, \bibinfo {author}
  {\bibfnamefont {P.~L.}\ \bibnamefont {Kuhns}}, \bibinfo {author}
  {\bibfnamefont {A.~P.}\ \bibnamefont {Reyes}},\ and\ \bibinfo {author}
  {\bibfnamefont {G.-q.}\ \bibnamefont {Zheng}},\ }\bibfield  {title} {\bibinfo
  {title} {Charge-density-wave order takes over antiferromagnetism in
  {Bi$_2$Sr$_{2-x}$La$_x$CuO$_6$} superconductors},\ }\href
  {https://doi.org/10.1038/s41467-017-01465-9} {\bibfield  {journal} {\bibinfo
  {journal} {Nature Communications}\ }\textbf {\bibinfo {volume} {8}},\
  \bibinfo {pages} {1267} (\bibinfo {year} {2017})}\BibitemShut {NoStop}%
\bibitem [{\citenamefont {Song}\ \emph {et~al.}(2021)\citenamefont {Song},
  \citenamefont {Ying}, \citenamefont {Chen}, \citenamefont {Han},
  \citenamefont {Wu}, \citenamefont {Schnyder}, \citenamefont {Huang},
  \citenamefont {Guo},\ and\ \citenamefont {Chen}}]{Song2021}%
  \BibitemOpen
  \bibfield  {author} {\bibinfo {author} {\bibfnamefont {Y.}~\bibnamefont
  {Song}}, \bibinfo {author} {\bibfnamefont {T.}~\bibnamefont {Ying}}, \bibinfo
  {author} {\bibfnamefont {X.}~\bibnamefont {Chen}}, \bibinfo {author}
  {\bibfnamefont {X.}~\bibnamefont {Han}}, \bibinfo {author} {\bibfnamefont
  {X.}~\bibnamefont {Wu}}, \bibinfo {author} {\bibfnamefont {A.~P.}\
  \bibnamefont {Schnyder}}, \bibinfo {author} {\bibfnamefont {Y.}~\bibnamefont
  {Huang}}, \bibinfo {author} {\bibfnamefont {J.-g.}\ \bibnamefont {Guo}},\
  and\ \bibinfo {author} {\bibfnamefont {X.}~\bibnamefont {Chen}},\ }\bibfield
  {title} {\bibinfo {title} {Competition of superconductivity and charge
  density wave in selective oxidized {CsV$_{3}$Sb$_{5}$} thin flakes},\ }\href
  {https://doi.org/10.1103/PhysRevLett.127.237001} {\bibfield  {journal}
  {\bibinfo  {journal} {Phys. Rev. Lett.}\ }\textbf {\bibinfo {volume} {127}},\
  \bibinfo {pages} {237001} (\bibinfo {year} {2021})}\BibitemShut {NoStop}%
\bibitem [{\citenamefont {Zeng}\ \emph {et~al.}(2022)\citenamefont {Zeng},
  \citenamefont {Hu}, \citenamefont {Wang}, \citenamefont {Sun}, \citenamefont
  {Yang}, \citenamefont {Boubeche}, \citenamefont {Luo}, \citenamefont {He},
  \citenamefont {Cheng}, \citenamefont {Yao},\ and\ \citenamefont
  {Luo}}]{Zeng2022}%
  \BibitemOpen
  \bibfield  {author} {\bibinfo {author} {\bibfnamefont {L.}~\bibnamefont
  {Zeng}}, \bibinfo {author} {\bibfnamefont {X.}~\bibnamefont {Hu}}, \bibinfo
  {author} {\bibfnamefont {N.}~\bibnamefont {Wang}}, \bibinfo {author}
  {\bibfnamefont {J.}~\bibnamefont {Sun}}, \bibinfo {author} {\bibfnamefont
  {P.}~\bibnamefont {Yang}}, \bibinfo {author} {\bibfnamefont {M.}~\bibnamefont
  {Boubeche}}, \bibinfo {author} {\bibfnamefont {S.}~\bibnamefont {Luo}},
  \bibinfo {author} {\bibfnamefont {Y.}~\bibnamefont {He}}, \bibinfo {author}
  {\bibfnamefont {J.}~\bibnamefont {Cheng}}, \bibinfo {author} {\bibfnamefont
  {D.-X.}\ \bibnamefont {Yao}},\ and\ \bibinfo {author} {\bibfnamefont
  {H.}~\bibnamefont {Luo}},\ }\bibfield  {title} {\bibinfo {title} {Interplay
  between charge-density-wave, superconductivity, and ferromagnetism in
  {CuIr$_{2–x}$Cr$_x$Te$_4$} chalcogenides},\ }\href
  {https://doi.org/10.1021/acs.jpclett.2c00404} {\bibfield  {journal} {\bibinfo
   {journal} {The Journal of Physical Chemistry Letters}\ }\textbf {\bibinfo
  {volume} {13}},\ \bibinfo {pages} {2442} (\bibinfo {year}
  {2022})}\BibitemShut {NoStop}%
\bibitem [{\citenamefont {Teng}\ \emph {et~al.}(2023)\citenamefont {Teng},
  \citenamefont {Oh}, \citenamefont {Tan}, \citenamefont {Chen}, \citenamefont
  {Huang}, \citenamefont {Gao}, \citenamefont {Yin}, \citenamefont {Chu},
  \citenamefont {Hashimoto}, \citenamefont {Lu}, \citenamefont {Jozwiak},
  \citenamefont {Bostwick}, \citenamefont {Rotenberg}, \citenamefont
  {Granroth}, \citenamefont {Yan}, \citenamefont {Birgeneau}, \citenamefont
  {Dai},\ and\ \citenamefont {Yi}}]{Teng2023}%
  \BibitemOpen
  \bibfield  {author} {\bibinfo {author} {\bibfnamefont {X.}~\bibnamefont
  {Teng}}, \bibinfo {author} {\bibfnamefont {J.~S.}\ \bibnamefont {Oh}},
  \bibinfo {author} {\bibfnamefont {H.}~\bibnamefont {Tan}}, \bibinfo {author}
  {\bibfnamefont {L.}~\bibnamefont {Chen}}, \bibinfo {author} {\bibfnamefont
  {J.}~\bibnamefont {Huang}}, \bibinfo {author} {\bibfnamefont
  {B.}~\bibnamefont {Gao}}, \bibinfo {author} {\bibfnamefont {J.-X.}\
  \bibnamefont {Yin}}, \bibinfo {author} {\bibfnamefont {J.-H.}\ \bibnamefont
  {Chu}}, \bibinfo {author} {\bibfnamefont {M.}~\bibnamefont {Hashimoto}},
  \bibinfo {author} {\bibfnamefont {D.}~\bibnamefont {Lu}}, \bibinfo {author}
  {\bibfnamefont {C.}~\bibnamefont {Jozwiak}}, \bibinfo {author} {\bibfnamefont
  {A.}~\bibnamefont {Bostwick}}, \bibinfo {author} {\bibfnamefont
  {E.}~\bibnamefont {Rotenberg}}, \bibinfo {author} {\bibfnamefont {G.~E.}\
  \bibnamefont {Granroth}}, \bibinfo {author} {\bibfnamefont {B.}~\bibnamefont
  {Yan}}, \bibinfo {author} {\bibfnamefont {R.~J.}\ \bibnamefont {Birgeneau}},
  \bibinfo {author} {\bibfnamefont {P.}~\bibnamefont {Dai}},\ and\ \bibinfo
  {author} {\bibfnamefont {M.}~\bibnamefont {Yi}},\ }\bibfield  {title}
  {\bibinfo {title} {Magnetism and charge density wave order in kagome
  {FeGe}},\ }\href {https://doi.org/10.1038/s41567-023-01985-w} {\bibfield
  {journal} {\bibinfo  {journal} {Nature Physics}\ }\textbf {\bibinfo {volume}
  {19}},\ \bibinfo {pages} {814} (\bibinfo {year} {2023})}\BibitemShut
  {NoStop}%
\bibitem [{\citenamefont {Ruderman}\ and\ \citenamefont
  {Kittel}(1954)}]{Ruderman1954}%
  \BibitemOpen
  \bibfield  {author} {\bibinfo {author} {\bibfnamefont {M.~A.}\ \bibnamefont
  {Ruderman}}\ and\ \bibinfo {author} {\bibfnamefont {C.}~\bibnamefont
  {Kittel}},\ }\bibfield  {title} {\bibinfo {title} {Indirect exchange coupling
  of nuclear magnetic moments by conduction electrons},\ }\href
  {https://doi.org/10.1103/PhysRev.96.99} {\bibfield  {journal} {\bibinfo
  {journal} {Phys. Rev.}\ }\textbf {\bibinfo {volume} {96}},\ \bibinfo {pages}
  {99} (\bibinfo {year} {1954})}\BibitemShut {NoStop}%
\bibitem [{\citenamefont {Kasuya}(1956)}]{Kasuya1956}%
  \BibitemOpen
  \bibfield  {author} {\bibinfo {author} {\bibfnamefont {T.}~\bibnamefont
  {Kasuya}},\ }\bibfield  {title} {\bibinfo {title} {{A Theory of Metallic
  Ferro- and Antiferromagnetism on Zener's Model}},\ }\href
  {https://doi.org/10.1143/PTP.16.45} {\bibfield  {journal} {\bibinfo
  {journal} {Progress of Theoretical Physics}\ }\textbf {\bibinfo {volume}
  {16}},\ \bibinfo {pages} {45} (\bibinfo {year} {1956})}\BibitemShut {NoStop}%
\bibitem [{\citenamefont {Yosida}(1957)}]{Yosida1957}%
  \BibitemOpen
  \bibfield  {author} {\bibinfo {author} {\bibfnamefont {K.}~\bibnamefont
  {Yosida}},\ }\bibfield  {title} {\bibinfo {title} {Magnetic properties of
  {Cu-Mn} alloys},\ }\href {https://doi.org/10.1103/PhysRev.106.893} {\bibfield
   {journal} {\bibinfo  {journal} {Phys. Rev.}\ }\textbf {\bibinfo {volume}
  {106}},\ \bibinfo {pages} {893} (\bibinfo {year} {1957})}\BibitemShut
  {NoStop}%
\bibitem [{\citenamefont {Peierls}(1955)}]{Peierls1955}%
  \BibitemOpen
  \bibfield  {author} {\bibinfo {author} {\bibfnamefont {R.~E.}\ \bibnamefont
  {Peierls}},\ }\href@noop {} {\emph {\bibinfo {title} {Quantum Theory of
  Solids}}}\ (\bibinfo  {publisher} {Oxford University Press},\ \bibinfo {year}
  {1955})\BibitemShut {NoStop}%
\bibitem [{\citenamefont {Fr{\"o}hlich}(1954)}]{Frohlich1954}%
  \BibitemOpen
  \bibfield  {author} {\bibinfo {author} {\bibfnamefont {H.}~\bibnamefont
  {Fr{\"o}hlich}},\ }\bibfield  {title} {\bibinfo {title} {On the theory of
  superconductivity: the one-dimensional case},\ }\href
  {https://doi.org/10.1098/rspa.1954.0116} {\bibfield  {journal} {\bibinfo
  {journal} {Proceedings of the Royal Society of London A: Mathematical,
  Physical and Engineering Sciences}\ }\textbf {\bibinfo {volume} {223}},\
  \bibinfo {pages} {296} (\bibinfo {year} {1954})}\BibitemShut {NoStop}%
\bibitem [{\citenamefont {Kurumaji}\ \emph {et~al.}(2019)\citenamefont
  {Kurumaji}, \citenamefont {Nakajima}, \citenamefont {Hirschberger},
  \citenamefont {Kikkawa}, \citenamefont {Yamasaki}, \citenamefont {Sagayama},
  \citenamefont {Nakao}, \citenamefont {Taguchi}, \citenamefont {Arima},\ and\
  \citenamefont {Tokura}}]{Kurumaji2019}%
  \BibitemOpen
  \bibfield  {author} {\bibinfo {author} {\bibfnamefont {T.}~\bibnamefont
  {Kurumaji}}, \bibinfo {author} {\bibfnamefont {T.}~\bibnamefont {Nakajima}},
  \bibinfo {author} {\bibfnamefont {M.}~\bibnamefont {Hirschberger}}, \bibinfo
  {author} {\bibfnamefont {A.}~\bibnamefont {Kikkawa}}, \bibinfo {author}
  {\bibfnamefont {Y.}~\bibnamefont {Yamasaki}}, \bibinfo {author}
  {\bibfnamefont {H.}~\bibnamefont {Sagayama}}, \bibinfo {author}
  {\bibfnamefont {H.}~\bibnamefont {Nakao}}, \bibinfo {author} {\bibfnamefont
  {Y.}~\bibnamefont {Taguchi}}, \bibinfo {author} {\bibfnamefont {T.-h.}\
  \bibnamefont {Arima}},\ and\ \bibinfo {author} {\bibfnamefont
  {Y.}~\bibnamefont {Tokura}},\ }\bibfield  {title} {\bibinfo {title}
  {{Skyrmion lattice with a giant topological Hall effect in a frustrated
  triangular-lattice magnet}},\ }\href@noop {} {\bibfield  {journal} {\bibinfo
  {journal} {Science}\ }\textbf {\bibinfo {volume} {365}},\ \bibinfo {pages}
  {914} (\bibinfo {year} {2019})}\BibitemShut {NoStop}%
\bibitem [{\citenamefont {Inosov}\ \emph {et~al.}(2009)\citenamefont {Inosov},
  \citenamefont {Evtushinsky}, \citenamefont {Koitzsch}, \citenamefont
  {Zabolotnyy}, \citenamefont {Borisenko}, \citenamefont {Kordyuk},
  \citenamefont {Frontzek}, \citenamefont {Loewenhaupt}, \citenamefont
  {L{\"o}ser}, \citenamefont {Mazilu}, \citenamefont {Bitterlich},
  \citenamefont {Behr}, \citenamefont {Hoffmann}, \citenamefont {Follath},\
  and\ \citenamefont {B{\"u}chner}}]{Inosov2009}%
  \BibitemOpen
  \bibfield  {author} {\bibinfo {author} {\bibfnamefont {D.~S.}\ \bibnamefont
  {Inosov}}, \bibinfo {author} {\bibfnamefont {D.~V.}\ \bibnamefont
  {Evtushinsky}}, \bibinfo {author} {\bibfnamefont {A.}~\bibnamefont
  {Koitzsch}}, \bibinfo {author} {\bibfnamefont {V.~B.}\ \bibnamefont
  {Zabolotnyy}}, \bibinfo {author} {\bibfnamefont {S.~V.}\ \bibnamefont
  {Borisenko}}, \bibinfo {author} {\bibfnamefont {A.~A.}\ \bibnamefont
  {Kordyuk}}, \bibinfo {author} {\bibfnamefont {M.}~\bibnamefont {Frontzek}},
  \bibinfo {author} {\bibfnamefont {M.}~\bibnamefont {Loewenhaupt}}, \bibinfo
  {author} {\bibfnamefont {W.}~\bibnamefont {L{\"o}ser}}, \bibinfo {author}
  {\bibfnamefont {I.}~\bibnamefont {Mazilu}}, \bibinfo {author} {\bibfnamefont
  {H.}~\bibnamefont {Bitterlich}}, \bibinfo {author} {\bibfnamefont
  {G.}~\bibnamefont {Behr}}, \bibinfo {author} {\bibfnamefont {J.-U.}\
  \bibnamefont {Hoffmann}}, \bibinfo {author} {\bibfnamefont {R.}~\bibnamefont
  {Follath}},\ and\ \bibinfo {author} {\bibfnamefont {B.}~\bibnamefont
  {B{\"u}chner}},\ }\bibfield  {title} {\bibinfo {title} {Electronic structure
  and nesting-driven enhancement of the {RKKY} interaction at the magnetic
  ordering propagation vector in {Gd$_{2}$PdSi$_{3}$} and
  {Tb$_{2}$PdSi$_{3}$}},\ }\href
  {https://doi.org/10.1103/PhysRevLett.102.046401} {\bibfield  {journal}
  {\bibinfo  {journal} {Phys. Rev. Lett.}\ }\textbf {\bibinfo {volume} {102}},\
  \bibinfo {pages} {046401} (\bibinfo {year} {2009})}\BibitemShut {NoStop}%
\bibitem [{\citenamefont {Salters}\ \emph {et~al.}(2023)\citenamefont
  {Salters}, \citenamefont {Orlandi}, \citenamefont {Berry}, \citenamefont
  {Khoury}, \citenamefont {Whittaker}, \citenamefont {Manuel},\ and\
  \citenamefont {Schoop}}]{Salters2023}%
  \BibitemOpen
  \bibfield  {author} {\bibinfo {author} {\bibfnamefont {T.~H.}\ \bibnamefont
  {Salters}}, \bibinfo {author} {\bibfnamefont {F.}~\bibnamefont {Orlandi}},
  \bibinfo {author} {\bibfnamefont {T.}~\bibnamefont {Berry}}, \bibinfo
  {author} {\bibfnamefont {J.~F.}\ \bibnamefont {Khoury}}, \bibinfo {author}
  {\bibfnamefont {E.}~\bibnamefont {Whittaker}}, \bibinfo {author}
  {\bibfnamefont {P.}~\bibnamefont {Manuel}},\ and\ \bibinfo {author}
  {\bibfnamefont {L.~M.}\ \bibnamefont {Schoop}},\ }\bibfield  {title}
  {\bibinfo {title} {Charge density wave-templated spin cycloid in topological
  semimetal {NdSb$_{x}$Te$_{\text{2-x-}\delta}$}},\ }\href
  {https://doi.org/10.1103/PhysRevMaterials.7.044203} {\bibfield  {journal}
  {\bibinfo  {journal} {Phys. Rev. Mater.}\ }\textbf {\bibinfo {volume} {7}},\
  \bibinfo {pages} {044203} (\bibinfo {year} {2023})}\BibitemShut {NoStop}%
\bibitem [{\citenamefont {Coelho}\ \emph {et~al.}(2019)\citenamefont {Coelho},
  \citenamefont {Nguyen~Cong}, \citenamefont {Bonilla}, \citenamefont
  {Kolekar}, \citenamefont {Phan}, \citenamefont {Avila}, \citenamefont
  {Asensio}, \citenamefont {Oleynik},\ and\ \citenamefont
  {Batzill}}]{Coelho2019}%
  \BibitemOpen
  \bibfield  {author} {\bibinfo {author} {\bibfnamefont {P.~M.}\ \bibnamefont
  {Coelho}}, \bibinfo {author} {\bibfnamefont {K.}~\bibnamefont {Nguyen~Cong}},
  \bibinfo {author} {\bibfnamefont {M.}~\bibnamefont {Bonilla}}, \bibinfo
  {author} {\bibfnamefont {S.}~\bibnamefont {Kolekar}}, \bibinfo {author}
  {\bibfnamefont {M.-H.}\ \bibnamefont {Phan}}, \bibinfo {author}
  {\bibfnamefont {J.}~\bibnamefont {Avila}}, \bibinfo {author} {\bibfnamefont
  {M.~C.}\ \bibnamefont {Asensio}}, \bibinfo {author} {\bibfnamefont {I.~I.}\
  \bibnamefont {Oleynik}},\ and\ \bibinfo {author} {\bibfnamefont
  {M.}~\bibnamefont {Batzill}},\ }\bibfield  {title} {\bibinfo {title} {Charge
  density wave state suppresses ferromagnetic ordering in {VSe$_2$}
  monolayers},\ }\href {https://doi.org/10.1021/acs.jpcc.9b04281} {\bibfield
  {journal} {\bibinfo  {journal} {The Journal of Physical Chemistry C}\
  }\textbf {\bibinfo {volume} {123}},\ \bibinfo {pages} {14089} (\bibinfo
  {year} {2019})}\BibitemShut {NoStop}%
\bibitem [{\citenamefont {Ramakrishnan}\ \emph {et~al.}(2020)\citenamefont
  {Ramakrishnan}, \citenamefont {Sch{\"o}nleber}, \citenamefont {Rekis},
  \citenamefont {van Well}, \citenamefont {Noohinejad}, \citenamefont {van
  Smaalen}, \citenamefont {Tolkiehn}, \citenamefont {Paulmann}, \citenamefont
  {Bag}, \citenamefont {Thamizhavel}, \citenamefont {Pal},\ and\ \citenamefont
  {Ramakrishnan}}]{Ramakrishnan2020}%
  \BibitemOpen
  \bibfield  {author} {\bibinfo {author} {\bibfnamefont {S.}~\bibnamefont
  {Ramakrishnan}}, \bibinfo {author} {\bibfnamefont {A.}~\bibnamefont
  {Sch{\"o}nleber}}, \bibinfo {author} {\bibfnamefont {T.}~\bibnamefont
  {Rekis}}, \bibinfo {author} {\bibfnamefont {N.}~\bibnamefont {van Well}},
  \bibinfo {author} {\bibfnamefont {L.}~\bibnamefont {Noohinejad}}, \bibinfo
  {author} {\bibfnamefont {S.}~\bibnamefont {van Smaalen}}, \bibinfo {author}
  {\bibfnamefont {M.}~\bibnamefont {Tolkiehn}}, \bibinfo {author}
  {\bibfnamefont {C.}~\bibnamefont {Paulmann}}, \bibinfo {author}
  {\bibfnamefont {B.}~\bibnamefont {Bag}}, \bibinfo {author} {\bibfnamefont
  {A.}~\bibnamefont {Thamizhavel}}, \bibinfo {author} {\bibfnamefont
  {D.}~\bibnamefont {Pal}},\ and\ \bibinfo {author} {\bibfnamefont
  {S.}~\bibnamefont {Ramakrishnan}},\ }\bibfield  {title} {\bibinfo {title}
  {Unusual charge density wave transition and absence of magnetic ordering in
  {Er$_2$Ir$_3$Si$_5$}},\ }\href {https://doi.org/10.1103/PhysRevB.101.060101}
  {\bibfield  {journal} {\bibinfo  {journal} {Phys. Rev. B}\ }\textbf {\bibinfo
  {volume} {101}},\ \bibinfo {pages} {060101} (\bibinfo {year}
  {2020})}\BibitemShut {NoStop}%
\bibitem [{\citenamefont {Zhou}\ \emph {et~al.}(2023)\citenamefont {Zhou},
  \citenamefont {Wang}, \citenamefont {Wang}, \citenamefont {Feng},
  \citenamefont {Yang},\ and\ \citenamefont {Shen}}]{Zhou2023}%
  \BibitemOpen
  \bibfield  {author} {\bibinfo {author} {\bibfnamefont {J.}~\bibnamefont
  {Zhou}}, \bibinfo {author} {\bibfnamefont {Z.}~\bibnamefont {Wang}}, \bibinfo
  {author} {\bibfnamefont {S.}~\bibnamefont {Wang}}, \bibinfo {author}
  {\bibfnamefont {Y.~P.}\ \bibnamefont {Feng}}, \bibinfo {author}
  {\bibfnamefont {M.}~\bibnamefont {Yang}},\ and\ \bibinfo {author}
  {\bibfnamefont {L.}~\bibnamefont {Shen}},\ }\bibfield  {title} {\bibinfo
  {title} {Coexistence of ferromagnetism and charge density waves in monolayer
  {LaBr$_2$}},\ }\href {https://doi.org/10.1039/D3NH00150D} {\bibfield
  {journal} {\bibinfo  {journal} {Nanoscale Horiz.}\ }\textbf {\bibinfo
  {volume} {8}},\ \bibinfo {pages} {1054} (\bibinfo {year} {2023})}\BibitemShut
  {NoStop}%
\bibitem [{\citenamefont {W{\"o}lfel}\ \emph {et~al.}(2010)\citenamefont
  {W{\"o}lfel}, \citenamefont {Li}, \citenamefont {Shimomura}, \citenamefont
  {Onodera},\ and\ \citenamefont {van Smaalen}}]{Wolfel2010}%
  \BibitemOpen
  \bibfield  {author} {\bibinfo {author} {\bibfnamefont {A.}~\bibnamefont
  {W{\"o}lfel}}, \bibinfo {author} {\bibfnamefont {L.}~\bibnamefont {Li}},
  \bibinfo {author} {\bibfnamefont {S.}~\bibnamefont {Shimomura}}, \bibinfo
  {author} {\bibfnamefont {H.}~\bibnamefont {Onodera}},\ and\ \bibinfo {author}
  {\bibfnamefont {S.}~\bibnamefont {van Smaalen}},\ }\bibfield  {title}
  {\bibinfo {title} {Commensurate charge-density wave with frustrated
  interchain coupling in {SmNiC$_2$}},\ }\href
  {https://doi.org/10.1103/PhysRevB.82.054120} {\bibfield  {journal} {\bibinfo
  {journal} {Physical Review B}\ }\textbf {\bibinfo {volume} {82}},\ \bibinfo
  {pages} {054120} (\bibinfo {year} {2010})}\BibitemShut {NoStop}%
\bibitem [{\citenamefont {Yamamoto}\ \emph {et~al.}(2013)\citenamefont
  {Yamamoto}, \citenamefont {Kondo}, \citenamefont {Maeda},\ and\ \citenamefont
  {Nogami}}]{Yamamoto2013}%
  \BibitemOpen
  \bibfield  {author} {\bibinfo {author} {\bibfnamefont {N.}~\bibnamefont
  {Yamamoto}}, \bibinfo {author} {\bibfnamefont {R.}~\bibnamefont {Kondo}},
  \bibinfo {author} {\bibfnamefont {H.}~\bibnamefont {Maeda}},\ and\ \bibinfo
  {author} {\bibfnamefont {Y.}~\bibnamefont {Nogami}},\ }\bibfield  {title}
  {\bibinfo {title} {Interplay of {Charge}-{Density} {Wave} and {Magnetic}
  {Order} in {Ternary} {Rare}-{Earth} {Nickel} {Carbides}, {RNiC$_2$} ({R}={Pr}
  and {Nd})},\ }\href {https://doi.org/10.7566/JPSJ.82.123701} {\bibfield
  {journal} {\bibinfo  {journal} {Journal of the Physical Society of Japan}\
  }\textbf {\bibinfo {volume} {82}},\ \bibinfo {pages} {123701} (\bibinfo
  {year} {2013})}\BibitemShut {NoStop}%
\bibitem [{\citenamefont {Shimomura}\ \emph {et~al.}(2016)\citenamefont
  {Shimomura}, \citenamefont {Hayashi}, \citenamefont {Hanasaki}, \citenamefont
  {Ohnuma}, \citenamefont {Kobayashi}, \citenamefont {Nakao}, \citenamefont
  {Mizumaki},\ and\ \citenamefont {Onodera}}]{Shimomura2016}%
  \BibitemOpen
  \bibfield  {author} {\bibinfo {author} {\bibfnamefont {S.}~\bibnamefont
  {Shimomura}}, \bibinfo {author} {\bibfnamefont {C.}~\bibnamefont {Hayashi}},
  \bibinfo {author} {\bibfnamefont {N.}~\bibnamefont {Hanasaki}}, \bibinfo
  {author} {\bibfnamefont {K.}~\bibnamefont {Ohnuma}}, \bibinfo {author}
  {\bibfnamefont {Y.}~\bibnamefont {Kobayashi}}, \bibinfo {author}
  {\bibfnamefont {H.}~\bibnamefont {Nakao}}, \bibinfo {author} {\bibfnamefont
  {M.}~\bibnamefont {Mizumaki}},\ and\ \bibinfo {author} {\bibfnamefont
  {H.}~\bibnamefont {Onodera}},\ }\bibfield  {title} {\bibinfo {title}
  {Multiple charge density wave transitions in the antiferromagnets {RNiC$_2$}
  {(R = Gd, Tb)}},\ }\href {https://doi.org/10.1103/PhysRevB.93.165108}
  {\bibfield  {journal} {\bibinfo  {journal} {Physical Review B}\ }\textbf
  {\bibinfo {volume} {93}},\ \bibinfo {pages} {165108} (\bibinfo {year}
  {2016})}\BibitemShut {NoStop}%
\bibitem [{\citenamefont {Maeda}\ \emph {et~al.}(2019)\citenamefont {Maeda},
  \citenamefont {Kondo},\ and\ \citenamefont {Nogami}}]{Maeda2019}%
  \BibitemOpen
  \bibfield  {author} {\bibinfo {author} {\bibfnamefont {H.}~\bibnamefont
  {Maeda}}, \bibinfo {author} {\bibfnamefont {R.}~\bibnamefont {Kondo}},\ and\
  \bibinfo {author} {\bibfnamefont {Y.}~\bibnamefont {Nogami}},\ }\bibfield
  {title} {\bibinfo {title} {Multiple charge density waves compete in ternary
  rare-earth nickel carbides, {RNiC$_2$ (R: Y, Dy, Ho, and Er)}},\ }\href
  {https://doi.org/10.1103/PhysRevB.100.104107} {\bibfield  {journal} {\bibinfo
   {journal} {Phys. Rev. B}\ }\textbf {\bibinfo {volume} {100}},\ \bibinfo
  {pages} {104107} (\bibinfo {year} {2019})}\BibitemShut {NoStop}%
\bibitem [{\citenamefont {Hanasaki}\ \emph {et~al.}(2017)\citenamefont
  {Hanasaki}, \citenamefont {Shimomura}, \citenamefont {Mikami}, \citenamefont
  {Nogami}, \citenamefont {Nakao},\ and\ \citenamefont
  {Onodera}}]{Hanasaki2017}%
  \BibitemOpen
  \bibfield  {author} {\bibinfo {author} {\bibfnamefont {N.}~\bibnamefont
  {Hanasaki}}, \bibinfo {author} {\bibfnamefont {S.}~\bibnamefont {Shimomura}},
  \bibinfo {author} {\bibfnamefont {K.}~\bibnamefont {Mikami}}, \bibinfo
  {author} {\bibfnamefont {Y.}~\bibnamefont {Nogami}}, \bibinfo {author}
  {\bibfnamefont {H.}~\bibnamefont {Nakao}},\ and\ \bibinfo {author}
  {\bibfnamefont {H.}~\bibnamefont {Onodera}},\ }\bibfield  {title} {\bibinfo
  {title} {Interplay between charge density wave and antiferromagnetic order in
  {GdNiC$_2$}},\ }\href {https://doi.org/10.1103/PhysRevB.95.085103} {\bibfield
   {journal} {\bibinfo  {journal} {Phys. Rev. B}\ }\textbf {\bibinfo {volume}
  {95}},\ \bibinfo {pages} {085103} (\bibinfo {year} {2017})}\BibitemShut
  {NoStop}%
\bibitem [{\citenamefont {Shimomura}\ \emph {et~al.}(2009)\citenamefont
  {Shimomura}, \citenamefont {Hayashi}, \citenamefont {Asaka}, \citenamefont
  {Wakabayashi}, \citenamefont {Mizumaki},\ and\ \citenamefont
  {Onodera}}]{Shimomura2009}%
  \BibitemOpen
  \bibfield  {author} {\bibinfo {author} {\bibfnamefont {S.}~\bibnamefont
  {Shimomura}}, \bibinfo {author} {\bibfnamefont {C.}~\bibnamefont {Hayashi}},
  \bibinfo {author} {\bibfnamefont {G.}~\bibnamefont {Asaka}}, \bibinfo
  {author} {\bibfnamefont {N.}~\bibnamefont {Wakabayashi}}, \bibinfo {author}
  {\bibfnamefont {M.}~\bibnamefont {Mizumaki}},\ and\ \bibinfo {author}
  {\bibfnamefont {H.}~\bibnamefont {Onodera}},\ }\bibfield  {title} {\bibinfo
  {title} {Charge-{Density}-{Wave} {Destruction} and {Ferromagnetic} {Order} in
  {SmNiC$_2$}},\ }\href {https://doi.org/10.1103/PhysRevLett.102.076404}
  {\bibfield  {journal} {\bibinfo  {journal} {Physical Review Letters}\
  }\textbf {\bibinfo {volume} {102}},\ \bibinfo {pages} {076404} (\bibinfo
  {year} {2009})}\BibitemShut {NoStop}%
\bibitem [{\citenamefont {Hanasaki}\ \emph {et~al.}(2012)\citenamefont
  {Hanasaki}, \citenamefont {Nogami}, \citenamefont {Kakinuma}, \citenamefont
  {Shimomura}, \citenamefont {Kosaka},\ and\ \citenamefont
  {Onodera}}]{Hanasaki2012}%
  \BibitemOpen
  \bibfield  {author} {\bibinfo {author} {\bibfnamefont {N.}~\bibnamefont
  {Hanasaki}}, \bibinfo {author} {\bibfnamefont {Y.}~\bibnamefont {Nogami}},
  \bibinfo {author} {\bibfnamefont {M.}~\bibnamefont {Kakinuma}}, \bibinfo
  {author} {\bibfnamefont {S.}~\bibnamefont {Shimomura}}, \bibinfo {author}
  {\bibfnamefont {M.}~\bibnamefont {Kosaka}},\ and\ \bibinfo {author}
  {\bibfnamefont {H.}~\bibnamefont {Onodera}},\ }\bibfield  {title} {\bibinfo
  {title} {Magnetic field switching of the charge-density-wave state in the
  lanthanide intermetallic {SmNiC$_2$}},\ }\href
  {https://doi.org/10.1103/PhysRevB.85.092402} {\bibfield  {journal} {\bibinfo
  {journal} {Physical Review B}\ }\textbf {\bibinfo {volume} {85}},\ \bibinfo
  {pages} {092402} (\bibinfo {year} {2012})}\BibitemShut {NoStop}%
\bibitem [{\citenamefont {Kolincio}\ \emph {et~al.}(2020)\citenamefont
  {Kolincio}, \citenamefont {Roman},\ and\ \citenamefont
  {Klimczuk}}]{Kolincio2020}%
  \BibitemOpen
  \bibfield  {author} {\bibinfo {author} {\bibfnamefont {K.~K.}\ \bibnamefont
  {Kolincio}}, \bibinfo {author} {\bibfnamefont {M.}~\bibnamefont {Roman}},\
  and\ \bibinfo {author} {\bibfnamefont {T.}~\bibnamefont {Klimczuk}},\
  }\bibfield  {title} {\bibinfo {title} {Enhanced mobility and large linear
  nonsaturating magnetoresistance in the magnetically ordered states of
  {TmNiC$_2$}},\ }\href {https://doi.org/10.1103/PhysRevLett.125.176601}
  {\bibfield  {journal} {\bibinfo  {journal} {Phys. Rev. Lett.}\ }\textbf
  {\bibinfo {volume} {125}},\ \bibinfo {pages} {176601} (\bibinfo {year}
  {2020})}\BibitemShut {NoStop}%
\bibitem [{\citenamefont {Kolincio}\ \emph {et~al.}(2024)\citenamefont
  {Kolincio}, \citenamefont {Roman}, \citenamefont {Garmroudi}, \citenamefont
  {Parzer}, \citenamefont {Bauer},\ and\ \citenamefont
  {Michor}}]{Kolincio2024}%
  \BibitemOpen
  \bibfield  {author} {\bibinfo {author} {\bibfnamefont {K.~K.}\ \bibnamefont
  {Kolincio}}, \bibinfo {author} {\bibfnamefont {M.}~\bibnamefont {Roman}},
  \bibinfo {author} {\bibfnamefont {F.}~\bibnamefont {Garmroudi}}, \bibinfo
  {author} {\bibfnamefont {M.}~\bibnamefont {Parzer}}, \bibinfo {author}
  {\bibfnamefont {E.}~\bibnamefont {Bauer}},\ and\ \bibinfo {author}
  {\bibfnamefont {H.}~\bibnamefont {Michor}},\ }\bibfield  {title} {\bibinfo
  {title} {Charge density wave, enhanced mobility, and large nonsaturating
  magnetoresistance across the magnetic states of {HoNiC$_{2}$} and
  {ErNiC$_{2}$}},\ }\href {https://doi.org/10.1103/PhysRevB.109.075154}
  {\bibfield  {journal} {\bibinfo  {journal} {Phys. Rev. B}\ }\textbf {\bibinfo
  {volume} {109}},\ \bibinfo {pages} {075154} (\bibinfo {year}
  {2024})}\BibitemShut {NoStop}%
\bibitem [{\citenamefont {Ray}\ \emph {et~al.}(2022)\citenamefont {Ray},
  \citenamefont {Sadhukhan}, \citenamefont {Richter}, \citenamefont {Facio},\
  and\ \citenamefont {van~den Brink}}]{Ray2022}%
  \BibitemOpen
  \bibfield  {author} {\bibinfo {author} {\bibfnamefont {R.}~\bibnamefont
  {Ray}}, \bibinfo {author} {\bibfnamefont {B.}~\bibnamefont {Sadhukhan}},
  \bibinfo {author} {\bibfnamefont {M.}~\bibnamefont {Richter}}, \bibinfo
  {author} {\bibfnamefont {J.~I.}\ \bibnamefont {Facio}},\ and\ \bibinfo
  {author} {\bibfnamefont {J.}~\bibnamefont {van~den Brink}},\ }\bibfield
  {title} {\bibinfo {title} {Tunable chirality of noncentrosymmetric magnetic
  {Weyl} semimetals in rare-earth carbides},\ }\href
  {https://doi.org/10.1038/s41535-022-00423-z} {\bibfield  {journal} {\bibinfo
  {journal} {npj Quantum Materials}\ }\textbf {\bibinfo {volume} {7}},\
  \bibinfo {pages} {19} (\bibinfo {year} {2022})}\BibitemShut {NoStop}%
\bibitem [{\citenamefont {Gr{\"u}ner}(1988)}]{Gruner1988}%
  \BibitemOpen
  \bibfield  {author} {\bibinfo {author} {\bibfnamefont {G.}~\bibnamefont
  {Gr{\"u}ner}},\ }\bibfield  {title} {\bibinfo {title} {The dynamics of
  charge-density waves},\ }\href {https://doi.org/10.1103/RevModPhys.60.1129}
  {\bibfield  {journal} {\bibinfo  {journal} {Rev. Mod. Phys.}\ }\textbf
  {\bibinfo {volume} {60}},\ \bibinfo {pages} {1129} (\bibinfo {year}
  {1988})}\BibitemShut {NoStop}%
\bibitem [{\citenamefont {Monceau}(2012)}]{Monceau2012}%
  \BibitemOpen
  \bibfield  {author} {\bibinfo {author} {\bibfnamefont {P.}~\bibnamefont
  {Monceau}},\ }\bibfield  {title} {\bibinfo {title} {Electronic crystals: an
  experimental overview},\ }\href
  {https://doi.org/10.1080/00018732.2012.719674} {\bibfield  {journal}
  {\bibinfo  {journal} {Advances in Physics}\ }\textbf {\bibinfo {volume}
  {61}},\ \bibinfo {pages} {325} (\bibinfo {year} {2012})}\BibitemShut
  {NoStop}%
\bibitem [{\citenamefont {Chang}\ \emph {et~al.}(2016)\citenamefont {Chang},
  \citenamefont {Blackburn}, \citenamefont {Ivashko}, \citenamefont {Holmes},
  \citenamefont {Christensen}, \citenamefont {H{\"u}cker}, \citenamefont
  {Liang}, \citenamefont {Bonn}, \citenamefont {Hardy}, \citenamefont
  {R{\"u}tt}, \citenamefont {Zimmermann}, \citenamefont {Forgan},\ and\
  \citenamefont {Hayden}}]{Chang2016}%
  \BibitemOpen
  \bibfield  {author} {\bibinfo {author} {\bibfnamefont {J.}~\bibnamefont
  {Chang}}, \bibinfo {author} {\bibfnamefont {E.}~\bibnamefont {Blackburn}},
  \bibinfo {author} {\bibfnamefont {O.}~\bibnamefont {Ivashko}}, \bibinfo
  {author} {\bibfnamefont {A.~T.}\ \bibnamefont {Holmes}}, \bibinfo {author}
  {\bibfnamefont {N.~B.}\ \bibnamefont {Christensen}}, \bibinfo {author}
  {\bibfnamefont {M.}~\bibnamefont {H{\"u}cker}}, \bibinfo {author}
  {\bibfnamefont {R.}~\bibnamefont {Liang}}, \bibinfo {author} {\bibfnamefont
  {D.~A.}\ \bibnamefont {Bonn}}, \bibinfo {author} {\bibfnamefont {W.~N.}\
  \bibnamefont {Hardy}}, \bibinfo {author} {\bibfnamefont {U.}~\bibnamefont
  {R{\"u}tt}}, \bibinfo {author} {\bibfnamefont {M.~v.}\ \bibnamefont
  {Zimmermann}}, \bibinfo {author} {\bibfnamefont {E.~M.}\ \bibnamefont
  {Forgan}},\ and\ \bibinfo {author} {\bibfnamefont {S.~M.}\ \bibnamefont
  {Hayden}},\ }\bibfield  {title} {\bibinfo {title} {Magnetic field controlled
  charge density wave coupling in underdoped {YBa$_2$Cu$_3$O$_{6+\text{x}}$}},\
  }\href {https://doi.org/10.1038/ncomms11494} {\bibfield  {journal} {\bibinfo
  {journal} {Nature Communications}\ }\textbf {\bibinfo {volume} {7}},\
  \bibinfo {pages} {11494} (\bibinfo {year} {2016})}\BibitemShut {NoStop}%
\bibitem [{\citenamefont {Roman}\ \emph {et~al.}(2023)\citenamefont {Roman},
  \citenamefont {Fritthum}, \citenamefont {St{\"o}ger}, \citenamefont
  {Adroja},\ and\ \citenamefont {Michor}}]{Roman2023}%
  \BibitemOpen
  \bibfield  {author} {\bibinfo {author} {\bibfnamefont {M.}~\bibnamefont
  {Roman}}, \bibinfo {author} {\bibfnamefont {M.}~\bibnamefont {Fritthum}},
  \bibinfo {author} {\bibfnamefont {B.}~\bibnamefont {St{\"o}ger}}, \bibinfo
  {author} {\bibfnamefont {D.~T.}\ \bibnamefont {Adroja}},\ and\ \bibinfo
  {author} {\bibfnamefont {H.}~\bibnamefont {Michor}},\ }\bibfield  {title}
  {\bibinfo {title} {Charge density wave and crystalline electric field effects
  in {TmNiC$_2$}},\ }\href {https://doi.org/10.1103/PhysRevB.107.125137}
  {\bibfield  {journal} {\bibinfo  {journal} {Phys. Rev. B}\ }\textbf {\bibinfo
  {volume} {107}},\ \bibinfo {pages} {125137} (\bibinfo {year}
  {2023})}\BibitemShut {NoStop}%
\bibitem [{Bruker()}]{bruker}%
  \BibitemOpen
  Bruker,\ \href@noop {} {\bibinfo {title} {{APEXII}, {SAINT} and {SADABS}}}
  (\bibinfo {year} {2023})\BibitemShut {NoStop}%
\bibitem [{\citenamefont {Strempfer}\ \emph {et~al.}(2013)\citenamefont
  {Strempfer}, \citenamefont {Francoual}, \citenamefont {Reuther},
  \citenamefont {Shukla}, \citenamefont {Skaugen}, \citenamefont
  {Schulte-Schrepping}, \citenamefont {Kracht},\ and\ \citenamefont
  {Franz}}]{DESY}%
  \BibitemOpen
  \bibfield  {author} {\bibinfo {author} {\bibfnamefont {J.}~\bibnamefont
  {Strempfer}}, \bibinfo {author} {\bibfnamefont {S.}~\bibnamefont
  {Francoual}}, \bibinfo {author} {\bibfnamefont {D.}~\bibnamefont {Reuther}},
  \bibinfo {author} {\bibfnamefont {D.~K.}\ \bibnamefont {Shukla}}, \bibinfo
  {author} {\bibfnamefont {A.}~\bibnamefont {Skaugen}}, \bibinfo {author}
  {\bibfnamefont {H.}~\bibnamefont {Schulte-Schrepping}}, \bibinfo {author}
  {\bibfnamefont {T.}~\bibnamefont {Kracht}},\ and\ \bibinfo {author}
  {\bibfnamefont {H.}~\bibnamefont {Franz}},\ }\bibfield  {title} {\bibinfo
  {title} {{Resonant scattering and diffraction beamline P09 at~PETRA III}},\
  }\href {https://doi.org/10.1107/S0909049513009011} {\bibfield  {journal}
  {\bibinfo  {journal} {Journal of Synchrotron Radiation}\ }\textbf {\bibinfo
  {volume} {20}},\ \bibinfo {pages} {541} (\bibinfo {year} {2013})}\BibitemShut
  {NoStop}%
\bibitem [{\citenamefont {Francoual}\ \emph {et~al.}(2015)\citenamefont
  {Francoual}, \citenamefont {Strempfer}, \citenamefont {Warren}, \citenamefont
  {Liu}, \citenamefont {Skaugen}, \citenamefont {Poli}, \citenamefont {Blume},
  \citenamefont {Wolff-Fabris}, \citenamefont {Canfield},\ and\ \citenamefont
  {Lograsso}}]{DESY_S}%
  \BibitemOpen
  \bibfield  {author} {\bibinfo {author} {\bibfnamefont {S.}~\bibnamefont
  {Francoual}}, \bibinfo {author} {\bibfnamefont {J.}~\bibnamefont
  {Strempfer}}, \bibinfo {author} {\bibfnamefont {J.}~\bibnamefont {Warren}},
  \bibinfo {author} {\bibfnamefont {Y.}~\bibnamefont {Liu}}, \bibinfo {author}
  {\bibfnamefont {A.}~\bibnamefont {Skaugen}}, \bibinfo {author} {\bibfnamefont
  {S.}~\bibnamefont {Poli}}, \bibinfo {author} {\bibfnamefont {J.}~\bibnamefont
  {Blume}}, \bibinfo {author} {\bibfnamefont {F.}~\bibnamefont {Wolff-Fabris}},
  \bibinfo {author} {\bibfnamefont {P.~C.}\ \bibnamefont {Canfield}},\ and\
  \bibinfo {author} {\bibfnamefont {T.}~\bibnamefont {Lograsso}},\ }\bibfield
  {title} {\bibinfo {title} {{Single-crystal X-ray diffraction and resonant
  X-ray magnetic scattering at helium-3 temperatures in high magnetic fields at
  beamline P09 at PETRA III}},\ }\href
  {https://doi.org/10.1107/S1600577515014149} {\bibfield  {journal} {\bibinfo
  {journal} {Journal of Synchrotron Radiation}\ }\textbf {\bibinfo {volume}
  {22}},\ \bibinfo {pages} {1207} (\bibinfo {year} {2015})}\BibitemShut
  {NoStop}%
\bibitem [{SM()}]{SM}%
  \BibitemOpen
  \href@noop {} {}\bibinfo {note} {See Supplemental Material at [URL will be
  inserted by publisher] further details and descriptions of the used methods,
  details of the temperature rise estimate and complementary data (see also
  references [36–41] therein)}\BibitemShut {NoStop}%
\bibitem [{\citenamefont {Kriminski}\ \emph {et~al.}(2003)\citenamefont
  {Kriminski}, \citenamefont {Kazmierczak},\ and\ \citenamefont
  {Thorne}}]{Kriminski_2003}%
  \BibitemOpen
  \bibfield  {author} {\bibinfo {author} {\bibfnamefont {S.}~\bibnamefont
  {Kriminski}}, \bibinfo {author} {\bibfnamefont {M.}~\bibnamefont
  {Kazmierczak}},\ and\ \bibinfo {author} {\bibfnamefont {R.~E.}\ \bibnamefont
  {Thorne}},\ }\bibfield  {title} {\bibinfo {title} {{Heat transfer from
  protein crystals: implications for flash-cooling and X-ray beam heating}},\
  }\href {https://doi.org/10.1107/S0907444903002713} {\bibfield  {journal}
  {\bibinfo  {journal} {Acta Crystallographica Section D}\ }\textbf {\bibinfo
  {volume} {59}},\ \bibinfo {pages} {697} (\bibinfo {year} {2003})}\BibitemShut
  {NoStop}%
\bibitem [{\citenamefont {Mhaisekar}\ \emph {et~al.}(2005)\citenamefont
  {Mhaisekar}, \citenamefont {Kazmierczak},\ and\ \citenamefont
  {Banerjee}}]{Mhaisekar2005}%
  \BibitemOpen
  \bibfield  {author} {\bibinfo {author} {\bibfnamefont {A.}~\bibnamefont
  {Mhaisekar}}, \bibinfo {author} {\bibfnamefont {M.~J.}\ \bibnamefont
  {Kazmierczak}},\ and\ \bibinfo {author} {\bibfnamefont {R.}~\bibnamefont
  {Banerjee}},\ }\bibfield  {title} {\bibinfo {title} {{Three-dimensional
  numerical analysis of convection and conduction cooling of spherical
  biocrystals with localized heating from synchrotron X-ray beams}},\ }\href
  {https://doi.org/10.1107/S0909049505003250} {\bibfield  {journal} {\bibinfo
  {journal} {Journal of Synchrotron Radiation}\ }\textbf {\bibinfo {volume}
  {12}},\ \bibinfo {pages} {318} (\bibinfo {year} {2005})}\BibitemShut
  {NoStop}%
\bibitem [{\citenamefont {Berger}\ \emph {et~al.}(1999)\citenamefont {Berger},
  \citenamefont {Hubbell}, \citenamefont {Seltzer}, \citenamefont {Coursey},\
  and\ \citenamefont {Zucker}}]{XCOM}%
  \BibitemOpen
  \bibfield  {author} {\bibinfo {author} {\bibfnamefont {M.}~\bibnamefont
  {Berger}}, \bibinfo {author} {\bibfnamefont {J.}~\bibnamefont {Hubbell}},
  \bibinfo {author} {\bibfnamefont {S.}~\bibnamefont {Seltzer}}, \bibinfo
  {author} {\bibfnamefont {J.}~\bibnamefont {Coursey}},\ and\ \bibinfo {author}
  {\bibfnamefont {D.}~\bibnamefont {Zucker}},\ }\href@noop {} {\bibinfo {title}
  {Xcom: Photon cross section database (version 1.2)}} (\bibinfo {year}
  {1999})\BibitemShut {NoStop}%
\bibitem [{\citenamefont {Kim}\ \emph {et~al.}(2012)\citenamefont {Kim},
  \citenamefont {Rhyee},\ and\ \citenamefont {Kwon}}]{Kim2012}%
  \BibitemOpen
  \bibfield  {author} {\bibinfo {author} {\bibfnamefont {J.~H.}\ \bibnamefont
  {Kim}}, \bibinfo {author} {\bibfnamefont {J.-S.}\ \bibnamefont {Rhyee}},\
  and\ \bibinfo {author} {\bibfnamefont {Y.~S.}\ \bibnamefont {Kwon}},\
  }\bibfield  {title} {\bibinfo {title} {Magnon gap formation and charge
  density wave effect on thermoelectric properties in the {SmNiC$_2$}
  compound},\ }\href {https://doi.org/10.1103/PhysRevB.86.235101} {\bibfield
  {journal} {\bibinfo  {journal} {Phys. Rev. B}\ }\textbf {\bibinfo {volume}
  {86}},\ \bibinfo {pages} {235101} (\bibinfo {year} {2012})}\BibitemShut
  {NoStop}%
\bibitem [{\citenamefont {Daudin}\ \emph {et~al.}(1984)\citenamefont {Daudin},
  \citenamefont {Mutscheller},\ and\ \citenamefont {Wagner}}]{Daudin_1984}%
  \BibitemOpen
  \bibfield  {author} {\bibinfo {author} {\bibfnamefont {B.}~\bibnamefont
  {Daudin}}, \bibinfo {author} {\bibfnamefont {W.}~\bibnamefont
  {Mutscheller}},\ and\ \bibinfo {author} {\bibfnamefont {M.}~\bibnamefont
  {Wagner}},\ }\bibfield  {title} {\bibinfo {title} {Heat conductivity of the
  cooperative vibronic system {TmVO$_4$}},\ }\href
  {https://doi.org/https://doi.org/10.1002/pssb.2221220108} {\bibfield
  {journal} {\bibinfo  {journal} {Physica Status Solidi (B)}\ }\textbf
  {\bibinfo {volume} {122}},\ \bibinfo {pages} {73} (\bibinfo {year}
  {1984})}\BibitemShut {NoStop}%
\bibitem [{\citenamefont {Ellis}\ and\ \citenamefont
  {Bernstein}(2008)}]{cbflib}%
  \BibitemOpen
  \bibfield  {author} {\bibinfo {author} {\bibfnamefont {P.~J.}\ \bibnamefont
  {Ellis}}\ and\ \bibinfo {author} {\bibfnamefont {H.~J.}\ \bibnamefont
  {Bernstein}},\ }\href@noop {} {\bibinfo {title} {{CBFlib}}} (\bibinfo {year}
  {2008})\BibitemShut {NoStop}%
\bibitem [{\citenamefont {Giannozzi}\ \emph {et~al.}(2009)\citenamefont
  {Giannozzi}, \citenamefont {Baroni}, \citenamefont {Bonini}, \citenamefont
  {Calandra}, \citenamefont {Car}, \citenamefont {Cavazzoni}, \citenamefont
  {Ceresoli}, \citenamefont {Chiarotti}, \citenamefont {Cococcioni},
  \citenamefont {Dabo}, \citenamefont {Dal~Corso}, \citenamefont
  {de~Gironcoli}, \citenamefont {Fabris}, \citenamefont {Fratesi},
  \citenamefont {Gebauer}, \citenamefont {Gerstmann}, \citenamefont
  {Gougoussis}, \citenamefont {Kokalj}, \citenamefont {Lazzeri}, \citenamefont
  {Martin-Samos}, \citenamefont {Marzari}, \citenamefont {Mauri}, \citenamefont
  {Mazzarello}, \citenamefont {Paolini}, \citenamefont {Pasquarello},
  \citenamefont {Paulatto}, \citenamefont {Sbraccia}, \citenamefont {Scandolo},
  \citenamefont {Sclauzero}, \citenamefont {Seitsonen}, \citenamefont
  {Smogunov}, \citenamefont {Umari},\ and\ \citenamefont
  {Wentzcovitch}}]{Giannozzi_2009}%
  \BibitemOpen
  \bibfield  {author} {\bibinfo {author} {\bibfnamefont {P.}~\bibnamefont
  {Giannozzi}}, \bibinfo {author} {\bibfnamefont {S.}~\bibnamefont {Baroni}},
  \bibinfo {author} {\bibfnamefont {N.}~\bibnamefont {Bonini}}, \bibinfo
  {author} {\bibfnamefont {M.}~\bibnamefont {Calandra}}, \bibinfo {author}
  {\bibfnamefont {R.}~\bibnamefont {Car}}, \bibinfo {author} {\bibfnamefont
  {C.}~\bibnamefont {Cavazzoni}}, \bibinfo {author} {\bibfnamefont
  {D.}~\bibnamefont {Ceresoli}}, \bibinfo {author} {\bibfnamefont {G.~L.}\
  \bibnamefont {Chiarotti}}, \bibinfo {author} {\bibfnamefont {M.}~\bibnamefont
  {Cococcioni}}, \bibinfo {author} {\bibfnamefont {I.}~\bibnamefont {Dabo}},
  \bibinfo {author} {\bibfnamefont {A.}~\bibnamefont {Dal~Corso}}, \bibinfo
  {author} {\bibfnamefont {S.}~\bibnamefont {de~Gironcoli}}, \bibinfo {author}
  {\bibfnamefont {S.}~\bibnamefont {Fabris}}, \bibinfo {author} {\bibfnamefont
  {G.}~\bibnamefont {Fratesi}}, \bibinfo {author} {\bibfnamefont
  {R.}~\bibnamefont {Gebauer}}, \bibinfo {author} {\bibfnamefont
  {U.}~\bibnamefont {Gerstmann}}, \bibinfo {author} {\bibfnamefont
  {C.}~\bibnamefont {Gougoussis}}, \bibinfo {author} {\bibfnamefont
  {A.}~\bibnamefont {Kokalj}}, \bibinfo {author} {\bibfnamefont
  {M.}~\bibnamefont {Lazzeri}}, \bibinfo {author} {\bibfnamefont
  {L.}~\bibnamefont {Martin-Samos}}, \bibinfo {author} {\bibfnamefont
  {N.}~\bibnamefont {Marzari}}, \bibinfo {author} {\bibfnamefont
  {F.}~\bibnamefont {Mauri}}, \bibinfo {author} {\bibfnamefont
  {R.}~\bibnamefont {Mazzarello}}, \bibinfo {author} {\bibfnamefont
  {S.}~\bibnamefont {Paolini}}, \bibinfo {author} {\bibfnamefont
  {A.}~\bibnamefont {Pasquarello}}, \bibinfo {author} {\bibfnamefont
  {L.}~\bibnamefont {Paulatto}}, \bibinfo {author} {\bibfnamefont
  {C.}~\bibnamefont {Sbraccia}}, \bibinfo {author} {\bibfnamefont
  {S.}~\bibnamefont {Scandolo}}, \bibinfo {author} {\bibfnamefont
  {G.}~\bibnamefont {Sclauzero}}, \bibinfo {author} {\bibfnamefont {A.~P.}\
  \bibnamefont {Seitsonen}}, \bibinfo {author} {\bibfnamefont {A.}~\bibnamefont
  {Smogunov}}, \bibinfo {author} {\bibfnamefont {P.}~\bibnamefont {Umari}},\
  and\ \bibinfo {author} {\bibfnamefont {R.~M.}\ \bibnamefont {Wentzcovitch}},\
  }\bibfield  {title} {\bibinfo {title} {Quantum espresso: a modular and
  open-source software project for quantum simulations of materials},\ }\href
  {https://doi.org/10.1088/0953-8984/21/39/395502} {\bibfield  {journal}
  {\bibinfo  {journal} {Journal of Physics: Condensed Matter}\ }\textbf
  {\bibinfo {volume} {21}},\ \bibinfo {pages} {395502} (\bibinfo {year}
  {2009})}\BibitemShut {NoStop}%
\bibitem [{\citenamefont {Giannozzi}\ \emph {et~al.}(2017)\citenamefont
  {Giannozzi}, \citenamefont {Andreussi}, \citenamefont {Brumme}, \citenamefont
  {Bunau}, \citenamefont {Nardelli}, \citenamefont {Calandra}, \citenamefont
  {Car}, \citenamefont {Cavazzoni}, \citenamefont {Ceresoli}, \citenamefont
  {Cococcioni} \emph {et~al.}}]{giannozzi2017advanced}%
  \BibitemOpen
  \bibfield  {author} {\bibinfo {author} {\bibfnamefont {P.}~\bibnamefont
  {Giannozzi}}, \bibinfo {author} {\bibfnamefont {O.}~\bibnamefont
  {Andreussi}}, \bibinfo {author} {\bibfnamefont {T.}~\bibnamefont {Brumme}},
  \bibinfo {author} {\bibfnamefont {O.}~\bibnamefont {Bunau}}, \bibinfo
  {author} {\bibfnamefont {M.~B.}\ \bibnamefont {Nardelli}}, \bibinfo {author}
  {\bibfnamefont {M.}~\bibnamefont {Calandra}}, \bibinfo {author}
  {\bibfnamefont {R.}~\bibnamefont {Car}}, \bibinfo {author} {\bibfnamefont
  {C.}~\bibnamefont {Cavazzoni}}, \bibinfo {author} {\bibfnamefont
  {D.}~\bibnamefont {Ceresoli}}, \bibinfo {author} {\bibfnamefont
  {M.}~\bibnamefont {Cococcioni}}, \emph {et~al.},\ }\bibfield  {title}
  {\bibinfo {title} {Advanced capabilities for materials modelling with quantum
  espresso},\ }\href@noop {} {\bibfield  {journal} {\bibinfo  {journal}
  {Journal of physics: Condensed matter}\ }\textbf {\bibinfo {volume} {29}},\
  \bibinfo {pages} {465901} (\bibinfo {year} {2017})}\BibitemShut {NoStop}%
\bibitem [{\citenamefont {Noffsinger}\ \emph {et~al.}(2010)\citenamefont
  {Noffsinger}, \citenamefont {Giustino}, \citenamefont {Malone}, \citenamefont
  {Park}, \citenamefont {Louie},\ and\ \citenamefont {Cohen}}]{epw1}%
  \BibitemOpen
  \bibfield  {author} {\bibinfo {author} {\bibfnamefont {J.}~\bibnamefont
  {Noffsinger}}, \bibinfo {author} {\bibfnamefont {F.}~\bibnamefont
  {Giustino}}, \bibinfo {author} {\bibfnamefont {B.~D.}\ \bibnamefont
  {Malone}}, \bibinfo {author} {\bibfnamefont {C.-H.}\ \bibnamefont {Park}},
  \bibinfo {author} {\bibfnamefont {S.~G.}\ \bibnamefont {Louie}},\ and\
  \bibinfo {author} {\bibfnamefont {M.~L.}\ \bibnamefont {Cohen}},\ }\bibfield
  {title} {\bibinfo {title} {Epw: A program for calculating the
  electron--phonon coupling using maximally localized wannier functions},\
  }\href@noop {} {\bibfield  {journal} {\bibinfo  {journal} {Computer Physics
  Communications}\ }\textbf {\bibinfo {volume} {181}},\ \bibinfo {pages} {2140}
  (\bibinfo {year} {2010})}\BibitemShut {NoStop}%
\bibitem [{\citenamefont {Ponc{\'e}}\ \emph {et~al.}(2016)\citenamefont
  {Ponc{\'e}}, \citenamefont {Margine}, \citenamefont {Verdi},\ and\
  \citenamefont {Giustino}}]{epw2}%
  \BibitemOpen
  \bibfield  {author} {\bibinfo {author} {\bibfnamefont {S.}~\bibnamefont
  {Ponc{\'e}}}, \bibinfo {author} {\bibfnamefont {E.~R.}\ \bibnamefont
  {Margine}}, \bibinfo {author} {\bibfnamefont {C.}~\bibnamefont {Verdi}},\
  and\ \bibinfo {author} {\bibfnamefont {F.}~\bibnamefont {Giustino}},\
  }\bibfield  {title} {\bibinfo {title} {Epw: Electron--phonon coupling,
  transport and superconducting properties using maximally localized wannier
  functions},\ }\href@noop {} {\bibfield  {journal} {\bibinfo  {journal}
  {Computer Physics Communications}\ }\textbf {\bibinfo {volume} {209}},\
  \bibinfo {pages} {116} (\bibinfo {year} {2016})}\BibitemShut {NoStop}%
\bibitem [{\citenamefont {Lee}\ \emph {et~al.}(2023)\citenamefont {Lee},
  \citenamefont {Ponc{\'e}}, \citenamefont {Bushick}, \citenamefont
  {Hajinazar}, \citenamefont {Lafuente-Bartolome}, \citenamefont {Leveillee},
  \citenamefont {Lian}, \citenamefont {Lihm}, \citenamefont {Macheda},
  \citenamefont {Mori} \emph {et~al.}}]{epw3}%
  \BibitemOpen
  \bibfield  {author} {\bibinfo {author} {\bibfnamefont {H.}~\bibnamefont
  {Lee}}, \bibinfo {author} {\bibfnamefont {S.}~\bibnamefont {Ponc{\'e}}},
  \bibinfo {author} {\bibfnamefont {K.}~\bibnamefont {Bushick}}, \bibinfo
  {author} {\bibfnamefont {S.}~\bibnamefont {Hajinazar}}, \bibinfo {author}
  {\bibfnamefont {J.}~\bibnamefont {Lafuente-Bartolome}}, \bibinfo {author}
  {\bibfnamefont {J.}~\bibnamefont {Leveillee}}, \bibinfo {author}
  {\bibfnamefont {C.}~\bibnamefont {Lian}}, \bibinfo {author} {\bibfnamefont
  {J.-M.}\ \bibnamefont {Lihm}}, \bibinfo {author} {\bibfnamefont
  {F.}~\bibnamefont {Macheda}}, \bibinfo {author} {\bibfnamefont
  {H.}~\bibnamefont {Mori}}, \emph {et~al.},\ }\bibfield  {title} {\bibinfo
  {title} {Electron--phonon physics from first principles using the epw code},\
  }\href@noop {} {\bibfield  {journal} {\bibinfo  {journal} {npj Computational
  Materials}\ }\textbf {\bibinfo {volume} {9}},\ \bibinfo {pages} {156}
  (\bibinfo {year} {2023})}\BibitemShut {NoStop}%
\bibitem [{\citenamefont {Hamann}(2013)}]{Hamann2013}%
  \BibitemOpen
  \bibfield  {author} {\bibinfo {author} {\bibfnamefont {D.~R.}\ \bibnamefont
  {Hamann}},\ }\bibfield  {title} {\bibinfo {title} {Optimized norm-conserving
  vanderbilt pseudopotentials},\ }\href
  {https://doi.org/10.1103/PhysRevB.88.085117} {\bibfield  {journal} {\bibinfo
  {journal} {Physical Review B}\ }\textbf {\bibinfo {volume} {88}},\ \bibinfo
  {pages} {085117} (\bibinfo {year} {2013})}\BibitemShut {NoStop}%
\bibitem [{\citenamefont {Perdew}\ \emph {et~al.}(1996)\citenamefont {Perdew},
  \citenamefont {Burke},\ and\ \citenamefont
  {Ernzerhof}}]{perdew1996generalized}%
  \BibitemOpen
  \bibfield  {author} {\bibinfo {author} {\bibfnamefont {J.~P.}\ \bibnamefont
  {Perdew}}, \bibinfo {author} {\bibfnamefont {K.}~\bibnamefont {Burke}},\ and\
  \bibinfo {author} {\bibfnamefont {M.}~\bibnamefont {Ernzerhof}},\ }\bibfield
  {title} {\bibinfo {title} {Generalized gradient approximation made simple},\
  }\href@noop {} {\bibfield  {journal} {\bibinfo  {journal} {Physical review
  letters}\ }\textbf {\bibinfo {volume} {77}},\ \bibinfo {pages} {3865}
  (\bibinfo {year} {1996})}\BibitemShut {NoStop}%
\bibitem [{\citenamefont {Koshikawa}\ \emph {et~al.}(1997)\citenamefont
  {Koshikawa}, \citenamefont {Onodera}, \citenamefont {Kosaka}, \citenamefont
  {Yamauchi}, \citenamefont {Ohashi},\ and\ \citenamefont
  {Yamaguchi}}]{Koshikawa1997}%
  \BibitemOpen
  \bibfield  {author} {\bibinfo {author} {\bibfnamefont {Y.}~\bibnamefont
  {Koshikawa}}, \bibinfo {author} {\bibfnamefont {H.}~\bibnamefont {Onodera}},
  \bibinfo {author} {\bibfnamefont {M.}~\bibnamefont {Kosaka}}, \bibinfo
  {author} {\bibfnamefont {H.}~\bibnamefont {Yamauchi}}, \bibinfo {author}
  {\bibfnamefont {M.}~\bibnamefont {Ohashi}},\ and\ \bibinfo {author}
  {\bibfnamefont {Y.}~\bibnamefont {Yamaguchi}},\ }\bibfield  {title} {\bibinfo
  {title} {Crystalline electric fields and magnetic properties of
  single-crystalline {RNiC$_2$} compounds {(R = Ho, Er and Tm)}},\ }\href
  {https://doi.org/https://doi.org/10.1016/S0304-8853(97)00177-7} {\bibfield
  {journal} {\bibinfo  {journal} {Journal of Magnetism and Magnetic Materials}\
  }\textbf {\bibinfo {volume} {173}},\ \bibinfo {pages} {72 } (\bibinfo {year}
  {1997})}\BibitemShut {NoStop}%
\bibitem [{\citenamefont {Aharoni}(1998)}]{Aharoni_1998}%
  \BibitemOpen
  \bibfield  {author} {\bibinfo {author} {\bibfnamefont {A.}~\bibnamefont
  {Aharoni}},\ }\bibfield  {title} {\bibinfo {title} {{Demagnetizing factors
  for rectangular ferromagnetic prisms}},\ }\href
  {https://doi.org/10.1063/1.367113} {\bibfield  {journal} {\bibinfo  {journal}
  {Journal of Applied Physics}\ }\textbf {\bibinfo {volume} {83}},\ \bibinfo
  {pages} {3432} (\bibinfo {year} {1998})}\BibitemShut {NoStop}%
\bibitem [{\citenamefont {McMillan}(1975)}]{McMillan1975}%
  \BibitemOpen
  \bibfield  {author} {\bibinfo {author} {\bibfnamefont {W.~L.}\ \bibnamefont
  {McMillan}},\ }\bibfield  {title} {\bibinfo {title} {Landau theory of
  charge-density waves in transition-metal dichalcogenides},\ }\href
  {https://doi.org/10.1103/PhysRevB.12.1187} {\bibfield  {journal} {\bibinfo
  {journal} {Phys. Rev. B}\ }\textbf {\bibinfo {volume} {12}},\ \bibinfo
  {pages} {1187} (\bibinfo {year} {1975})}\BibitemShut {NoStop}%
\bibitem [{\citenamefont {Friend}\ and\ \citenamefont
  {Jerome}(1979)}]{Friend_1979}%
  \BibitemOpen
  \bibfield  {author} {\bibinfo {author} {\bibfnamefont {R.~H.}\ \bibnamefont
  {Friend}}\ and\ \bibinfo {author} {\bibfnamefont {D.}~\bibnamefont
  {Jerome}},\ }\bibfield  {title} {\bibinfo {title} {Periodic lattice
  distortions and charge density waves in one- and two-dimensional metals},\
  }\href {https://doi.org/10.1088/0022-3719/12/8/009} {\bibfield  {journal}
  {\bibinfo  {journal} {Journal of Physics C: Solid State Physics}\ }\textbf
  {\bibinfo {volume} {12}},\ \bibinfo {pages} {1441} (\bibinfo {year}
  {1979})}\BibitemShut {NoStop}%
\bibitem [{\citenamefont {Varma}\ and\ \citenamefont
  {Simons}(1983)}]{Varma_1983}%
  \BibitemOpen
  \bibfield  {author} {\bibinfo {author} {\bibfnamefont {C.~M.}\ \bibnamefont
  {Varma}}\ and\ \bibinfo {author} {\bibfnamefont {A.~L.}\ \bibnamefont
  {Simons}},\ }\bibfield  {title} {\bibinfo {title} {Strong-coupling theory of
  charge-density-wave transitions},\ }\href
  {https://doi.org/10.1103/PhysRevLett.51.138} {\bibfield  {journal} {\bibinfo
  {journal} {Phys. Rev. Lett.}\ }\textbf {\bibinfo {volume} {51}},\ \bibinfo
  {pages} {138} (\bibinfo {year} {1983})}\BibitemShut {NoStop}%
\bibitem [{\citenamefont {Galli}\ \emph {et~al.}(2002)\citenamefont {Galli},
  \citenamefont {Feyerherm}, \citenamefont {Hendrikx}, \citenamefont {Dudzik},
  \citenamefont {Nieuwenhuys}, \citenamefont {Ramakrishnan}, \citenamefont
  {Brown}, \citenamefont {van Smaalen},\ and\ \citenamefont
  {Mydosh}}]{Galli2002}%
  \BibitemOpen
  \bibfield  {author} {\bibinfo {author} {\bibfnamefont {F.}~\bibnamefont
  {Galli}}, \bibinfo {author} {\bibfnamefont {R.}~\bibnamefont {Feyerherm}},
  \bibinfo {author} {\bibfnamefont {R.~W.~A.}\ \bibnamefont {Hendrikx}},
  \bibinfo {author} {\bibfnamefont {E.}~\bibnamefont {Dudzik}}, \bibinfo
  {author} {\bibfnamefont {G.~J.}\ \bibnamefont {Nieuwenhuys}}, \bibinfo
  {author} {\bibfnamefont {S.}~\bibnamefont {Ramakrishnan}}, \bibinfo {author}
  {\bibfnamefont {S.~D.}\ \bibnamefont {Brown}}, \bibinfo {author}
  {\bibfnamefont {S.}~\bibnamefont {van Smaalen}},\ and\ \bibinfo {author}
  {\bibfnamefont {J.~A.}\ \bibnamefont {Mydosh}},\ }\bibfield  {title}
  {\bibinfo {title} {Coexistence of charge density wave and antiferromagnetism
  in {Er$_5$Ir$_4$Si$_{10}$}},\ }\href
  {http://stacks.iop.org/0953-8984/14/i=20/a=302} {\bibfield  {journal}
  {\bibinfo  {journal} {Journal of Physics: Condensed Matter}\ }\textbf
  {\bibinfo {volume} {14}},\ \bibinfo {pages} {5067} (\bibinfo {year}
  {2002})}\BibitemShut {NoStop}%
\bibitem [{\citenamefont {Johannes}\ and\ \citenamefont
  {Mazin}(2008)}]{Johannes2008}%
  \BibitemOpen
  \bibfield  {author} {\bibinfo {author} {\bibfnamefont {M.~D.}\ \bibnamefont
  {Johannes}}\ and\ \bibinfo {author} {\bibfnamefont {I.~I.}\ \bibnamefont
  {Mazin}},\ }\bibfield  {title} {\bibinfo {title} {Fermi surface nesting and
  the origin of charge density waves in metals},\ }\href
  {https://doi.org/10.1103/PhysRevB.77.165135} {\bibfield  {journal} {\bibinfo
  {journal} {Phys. Rev. B}\ }\textbf {\bibinfo {volume} {77}},\ \bibinfo
  {pages} {165135} (\bibinfo {year} {2008})}\BibitemShut {NoStop}%
\bibitem [{\citenamefont {Lian}\ \emph {et~al.}(2019)\citenamefont {Lian},
  \citenamefont {Heil}, \citenamefont {Liu}, \citenamefont {Si}, \citenamefont
  {Giustino},\ and\ \citenamefont {Duan}}]{lian2019coexistence}%
  \BibitemOpen
  \bibfield  {author} {\bibinfo {author} {\bibfnamefont {C.-S.}\ \bibnamefont
  {Lian}}, \bibinfo {author} {\bibfnamefont {C.}~\bibnamefont {Heil}}, \bibinfo
  {author} {\bibfnamefont {X.}~\bibnamefont {Liu}}, \bibinfo {author}
  {\bibfnamefont {C.}~\bibnamefont {Si}}, \bibinfo {author} {\bibfnamefont
  {F.}~\bibnamefont {Giustino}},\ and\ \bibinfo {author} {\bibfnamefont
  {W.}~\bibnamefont {Duan}},\ }\bibfield  {title} {\bibinfo {title}
  {Coexistence of superconductivity with enhanced charge density wave order in
  the two-dimensional limit of tase2},\ }\href@noop {} {\bibfield  {journal}
  {\bibinfo  {journal} {The Journal of Physical Chemistry Letters}\ }\textbf
  {\bibinfo {volume} {10}},\ \bibinfo {pages} {4076} (\bibinfo {year}
  {2019})}\BibitemShut {NoStop}%
\bibitem [{\citenamefont {Roman}\ \emph {et~al.}(2025)\citenamefont {Roman},
  \citenamefont {Di~Cataldo}, \citenamefont {St\"oger}, \citenamefont
  {Reisinger}, \citenamefont {Morineau}, \citenamefont {Kolincio},\ and\
  \citenamefont {Michor}}]{Roman_2024_Y}%
  \BibitemOpen
  \bibfield  {author} {\bibinfo {author} {\bibfnamefont {M.}~\bibnamefont
  {Roman}}, \bibinfo {author} {\bibfnamefont {S.}~\bibnamefont {Di~Cataldo}},
  \bibinfo {author} {\bibfnamefont {B.}~\bibnamefont {St\"oger}}, \bibinfo
  {author} {\bibfnamefont {L.}~\bibnamefont {Reisinger}}, \bibinfo {author}
  {\bibfnamefont {E.}~\bibnamefont {Morineau}}, \bibinfo {author}
  {\bibfnamefont {K.~K.}\ \bibnamefont {Kolincio}},\ and\ \bibinfo {author}
  {\bibfnamefont {H.}~\bibnamefont {Michor}},\ }\bibfield  {title} {\bibinfo
  {title} {Competing charge density wave phases in ${\mathrm{ynic}}_{2}$},\
  }\href {https://doi.org/10.1103/PhysRevB.111.195101} {\bibfield  {journal}
  {\bibinfo  {journal} {Phys. Rev. B}\ }\textbf {\bibinfo {volume} {111}},\
  \bibinfo {pages} {195101} (\bibinfo {year} {2025})}\BibitemShut {NoStop}%
\bibitem [{\citenamefont {Steiner}\ \emph {et~al.}(2018)\citenamefont
  {Steiner}, \citenamefont {Michor}, \citenamefont {Sologub}, \citenamefont
  {Hinterleitner}, \citenamefont {H{\"o}fenstock}, \citenamefont {Waas},
  \citenamefont {Bauer}, \citenamefont {St{\"o}ger}, \citenamefont
  {Babizhetskyy}, \citenamefont {Levytskyy},\ and\ \citenamefont
  {Kotur}}]{Steiner2018}%
  \BibitemOpen
  \bibfield  {author} {\bibinfo {author} {\bibfnamefont {S.}~\bibnamefont
  {Steiner}}, \bibinfo {author} {\bibfnamefont {H.}~\bibnamefont {Michor}},
  \bibinfo {author} {\bibfnamefont {O.}~\bibnamefont {Sologub}}, \bibinfo
  {author} {\bibfnamefont {B.}~\bibnamefont {Hinterleitner}}, \bibinfo {author}
  {\bibfnamefont {F.}~\bibnamefont {H{\"o}fenstock}}, \bibinfo {author}
  {\bibfnamefont {M.}~\bibnamefont {Waas}}, \bibinfo {author} {\bibfnamefont
  {E.}~\bibnamefont {Bauer}}, \bibinfo {author} {\bibfnamefont
  {B.}~\bibnamefont {St{\"o}ger}}, \bibinfo {author} {\bibfnamefont
  {V.}~\bibnamefont {Babizhetskyy}}, \bibinfo {author} {\bibfnamefont
  {V.}~\bibnamefont {Levytskyy}},\ and\ \bibinfo {author} {\bibfnamefont
  {B.}~\bibnamefont {Kotur}},\ }\bibfield  {title} {\bibinfo {title}
  {Single-crystal study of the charge density wave metal {LuNiC$_2$}},\ }\href
  {https://doi.org/10.1103/PhysRevB.97.205115} {\bibfield  {journal} {\bibinfo
  {journal} {Phys. Rev. B}\ }\textbf {\bibinfo {volume} {97}},\ \bibinfo
  {pages} {205115} (\bibinfo {year} {2018})}\BibitemShut {NoStop}%
\bibitem [{\citenamefont {Laverock}\ \emph {et~al.}(2009)\citenamefont
  {Laverock}, \citenamefont {Haynes}, \citenamefont {Utfeld},\ and\
  \citenamefont {Dugdale}}]{Laverock2009}%
  \BibitemOpen
  \bibfield  {author} {\bibinfo {author} {\bibfnamefont {J.}~\bibnamefont
  {Laverock}}, \bibinfo {author} {\bibfnamefont {T.~D.}\ \bibnamefont
  {Haynes}}, \bibinfo {author} {\bibfnamefont {C.}~\bibnamefont {Utfeld}},\
  and\ \bibinfo {author} {\bibfnamefont {S.~B.}\ \bibnamefont {Dugdale}},\
  }\bibfield  {title} {\bibinfo {title} {Electronic structure of
  {$R{\text{NiC}}_{2}$ ($R=\text{Sm}$, Gd, and Nd)} intermetallic compounds},\
  }\href {https://doi.org/10.1103/PhysRevB.80.125111} {\bibfield  {journal}
  {\bibinfo  {journal} {Phys. Rev. B}\ }\textbf {\bibinfo {volume} {80}},\
  \bibinfo {pages} {125111} (\bibinfo {year} {2009})}\BibitemShut {NoStop}%
\bibitem [{\citenamefont {Kim}\ \emph {et~al.}(2013)\citenamefont {Kim},
  \citenamefont {Lee},\ and\ \citenamefont {Shim}}]{Kim_2013}%
  \BibitemOpen
  \bibfield  {author} {\bibinfo {author} {\bibfnamefont {J.~N.}\ \bibnamefont
  {Kim}}, \bibinfo {author} {\bibfnamefont {C.}~\bibnamefont {Lee}},\ and\
  \bibinfo {author} {\bibfnamefont {J.-H.}\ \bibnamefont {Shim}},\ }\bibfield
  {title} {\bibinfo {title} {Chemical and hydrostatic pressure effect on charge
  density waves of {SmNiC$_2$}},\ }\href
  {https://doi.org/10.1088/1367-2630/15/12/123018} {\bibfield  {journal}
  {\bibinfo  {journal} {New Journal of Physics}\ }\textbf {\bibinfo {volume}
  {15}},\ \bibinfo {pages} {123018} (\bibinfo {year} {2013})}\BibitemShut
  {NoStop}%
\bibitem [{\citenamefont {Hase}\ and\ \citenamefont
  {Yanagisawa}(2009)}]{Hase2009}%
  \BibitemOpen
  \bibfield  {author} {\bibinfo {author} {\bibfnamefont {I.}~\bibnamefont
  {Hase}}\ and\ \bibinfo {author} {\bibfnamefont {T.}~\bibnamefont
  {Yanagisawa}},\ }\bibfield  {title} {\bibinfo {title} {Electronic structure
  of {RNiC$_2$ (R = La, Y, and Th)}},\ }\href
  {https://doi.org/10.1143/JPSJ.78.084724} {\bibfield  {journal} {\bibinfo
  {journal} {Journal of the Physical Society of Japan}\ }\textbf {\bibinfo
  {volume} {78}},\ \bibinfo {pages} {084724} (\bibinfo {year}
  {2009})}\BibitemShut {NoStop}%
\bibitem [{\citenamefont {Hanasaki}\ \emph {et~al.}(2011)\citenamefont
  {Hanasaki}, \citenamefont {Mikami}, \citenamefont {Torigoe}, \citenamefont
  {Nogami}, \citenamefont {Shimomura}, \citenamefont {Kosaka},\ and\
  \citenamefont {Onodera}}]{Hanasaki_2011}%
  \BibitemOpen
  \bibfield  {author} {\bibinfo {author} {\bibfnamefont {N.}~\bibnamefont
  {Hanasaki}}, \bibinfo {author} {\bibfnamefont {K.}~\bibnamefont {Mikami}},
  \bibinfo {author} {\bibfnamefont {S.}~\bibnamefont {Torigoe}}, \bibinfo
  {author} {\bibfnamefont {Y.}~\bibnamefont {Nogami}}, \bibinfo {author}
  {\bibfnamefont {S.}~\bibnamefont {Shimomura}}, \bibinfo {author}
  {\bibfnamefont {M.}~\bibnamefont {Kosaka}},\ and\ \bibinfo {author}
  {\bibfnamefont {H.}~\bibnamefont {Onodera}},\ }\bibfield  {title} {\bibinfo
  {title} {Successive transition in rare-earth intermetallic compound
  {GdNiC$_2$}},\ }\href {https://doi.org/10.1088/1742-6596/320/1/012072}
  {\bibfield  {journal} {\bibinfo  {journal} {Journal of Physics: Conference
  Series}\ }\textbf {\bibinfo {volume} {320}},\ \bibinfo {pages} {012072}
  (\bibinfo {year} {2011})}\BibitemShut {NoStop}%
\bibitem [{\citenamefont {Onodera}\ \emph {et~al.}(1998)\citenamefont
  {Onodera}, \citenamefont {Koshikawa}, \citenamefont {Kosaka}, \citenamefont
  {Ohashi}, \citenamefont {Yamauchi},\ and\ \citenamefont
  {Yamaguchi}}]{Onodera1998}%
  \BibitemOpen
  \bibfield  {author} {\bibinfo {author} {\bibfnamefont {H.}~\bibnamefont
  {Onodera}}, \bibinfo {author} {\bibfnamefont {Y.}~\bibnamefont {Koshikawa}},
  \bibinfo {author} {\bibfnamefont {M.}~\bibnamefont {Kosaka}}, \bibinfo
  {author} {\bibfnamefont {M.}~\bibnamefont {Ohashi}}, \bibinfo {author}
  {\bibfnamefont {H.}~\bibnamefont {Yamauchi}},\ and\ \bibinfo {author}
  {\bibfnamefont {Y.}~\bibnamefont {Yamaguchi}},\ }\bibfield  {title} {\bibinfo
  {title} {Magnetic properties of single-crystalline {RNiC$_2$} compounds {(R =
  Ce, Pr, Nd and Sm)}},\ }\href
  {https://doi.org/http://dx.doi.org/10.1016/S0304-8853(97)01011-1} {\bibfield
  {journal} {\bibinfo  {journal} {Journal of Magnetism and Magnetic Materials}\
  }\textbf {\bibinfo {volume} {182}},\ \bibinfo {pages} {161 } (\bibinfo {year}
  {1998})}\BibitemShut {NoStop}%
\bibitem [{\citenamefont {Eiter}\ \emph {et~al.}(2013)\citenamefont {Eiter},
  \citenamefont {Lavagnini}, \citenamefont {Hackl}, \citenamefont {Nowadnick},
  \citenamefont {Kemper}, \citenamefont {Devereaux}, \citenamefont {Chu},
  \citenamefont {Analytis}, \citenamefont {Fisher},\ and\ \citenamefont
  {Degiorgi}}]{Eiter_2013}%
  \BibitemOpen
  \bibfield  {author} {\bibinfo {author} {\bibfnamefont {H.-M.}\ \bibnamefont
  {Eiter}}, \bibinfo {author} {\bibfnamefont {M.}~\bibnamefont {Lavagnini}},
  \bibinfo {author} {\bibfnamefont {R.}~\bibnamefont {Hackl}}, \bibinfo
  {author} {\bibfnamefont {E.~A.}\ \bibnamefont {Nowadnick}}, \bibinfo {author}
  {\bibfnamefont {A.~F.}\ \bibnamefont {Kemper}}, \bibinfo {author}
  {\bibfnamefont {T.~P.}\ \bibnamefont {Devereaux}}, \bibinfo {author}
  {\bibfnamefont {J.-H.}\ \bibnamefont {Chu}}, \bibinfo {author} {\bibfnamefont
  {J.~G.}\ \bibnamefont {Analytis}}, \bibinfo {author} {\bibfnamefont {I.~R.}\
  \bibnamefont {Fisher}},\ and\ \bibinfo {author} {\bibfnamefont
  {L.}~\bibnamefont {Degiorgi}},\ }\bibfield  {title} {\bibinfo {title}
  {Alternative route to charge density wave formation in multiband systems},\
  }\href {https://doi.org/10.1073/pnas.1214745110} {\bibfield  {journal}
  {\bibinfo  {journal} {Proceedings of the National Academy of Sciences}\
  }\textbf {\bibinfo {volume} {110}},\ \bibinfo {pages} {64} (\bibinfo {year}
  {2013})}\BibitemShut {NoStop}%
\bibitem [{\citenamefont {Zhu}\ \emph {et~al.}(2015)\citenamefont {Zhu},
  \citenamefont {Cao}, \citenamefont {Zhang}, \citenamefont {Plummer},\ and\
  \citenamefont {Guo}}]{Zhu2015}%
  \BibitemOpen
  \bibfield  {author} {\bibinfo {author} {\bibfnamefont {X.}~\bibnamefont
  {Zhu}}, \bibinfo {author} {\bibfnamefont {Y.}~\bibnamefont {Cao}}, \bibinfo
  {author} {\bibfnamefont {J.}~\bibnamefont {Zhang}}, \bibinfo {author}
  {\bibfnamefont {E.~W.}\ \bibnamefont {Plummer}},\ and\ \bibinfo {author}
  {\bibfnamefont {J.}~\bibnamefont {Guo}},\ }\bibfield  {title} {\bibinfo
  {title} {Classification of charge density waves based on their nature},\
  }\href {https://doi.org/10.1073/pnas.1424791112} {\bibfield  {journal}
  {\bibinfo  {journal} {Proceedings of the National Academy of Sciences}\
  }\textbf {\bibinfo {volume} {112}},\ \bibinfo {pages} {2367} (\bibinfo {year}
  {2015})}\BibitemShut {NoStop}%
\bibitem [{\citenamefont {Xuetao~Zhu}\ and\ \citenamefont
  {Plummer}(2017)}]{Zhu_2017}%
  \BibitemOpen
  \bibfield  {author} {\bibinfo {author} {\bibfnamefont {J.~Z.}\ \bibnamefont
  {Xuetao~Zhu}, \bibfnamefont {Jiandong~Guo}}\ and\ \bibinfo {author}
  {\bibfnamefont {E.~W.}\ \bibnamefont {Plummer}},\ }\bibfield  {title}
  {\bibinfo {title} {Misconceptions associated with the origin of charge
  density waves},\ }\href {https://doi.org/10.1080/23746149.2017.1343098}
  {\bibfield  {journal} {\bibinfo  {journal} {Advances in Physics: X}\ }\textbf
  {\bibinfo {volume} {2}},\ \bibinfo {pages} {622} (\bibinfo {year}
  {2017})}\BibitemShut {NoStop}%
\end{thebibliography}
 
%
\end{document}

\title{Supplemental Material for: \\ Coexistence of charge density wave and field-tuned magnetic states in TmNiC$_2$}

\author{Kamil K. Kolincio}

\affiliation{Faculty of Applied Physics and Mathematics, Gdansk University of Technology,
Narutowicza 11/12, 80-233 Gdansk, Poland}
\email{corresponding author: kamil.kolincio@pg.edu.pl}

\author{Marta Roman}

\affiliation{Institute of Solid State Physics, TU Wien, Wiedner Hauptstrasse 8–10, A-1040 Wien, Austria}

\affiliation{Institute of Physics and Applied Computer Science, Faculty of Applied Physics and Mathematics, Gdansk University of Technology, Narutowicza 11/12, 80-233 Gdansk, Poland}

\author{Valerio Tammurello}
\affiliation{Dipartimento di Fisica, Sapienza University of Rome, Piazzale Aldo Moro 5, 00185 Rome, Italy}

\author{Simone Di Cataldo}
\affiliation{Dipartimento di Fisica, Sapienza University of Rome, Piazzale Aldo Moro 5, 00185 Rome, Italy}

\author{Daniel Matulka}
\affiliation{X-Ray Center, TU Wien, Getreidemarkt 9, A-1060 Wien, Austria}

\author{Sonia Francoual}
\affiliation{Deutsches Elektronen-Synchrotron DESY, 22607 Hamburg, Germany}

\author{Berthold St\"{o}ger}
\affiliation{X-Ray Center, TU Wien, Getreidemarkt 9, A-1060 Wien, Austria}

\author{Herwig Michor} 
\affiliation{Institute of Solid State Physics, TU Wien, Wiedner Hauptstrasse 8–10, A-1040 Wien, Austria}

\maketitle
\section{Complementary experimental details}
\subsection{Demagnetization correction}

The magnetic field in all the field dependencies throughout the paper, and in the magnetic phase diagram, have been corrected for demagnetization factor based on relevant sample shape. The internal magnetic field $B$ is calculated using the formula:

\begin{equation}
B = \mu_0H-NM,
\end{equation}

Where $\mu_0H$ is the externally applied field, $N$ is the demagnetization factor depending on sample shape and $M$ is the magnetization.

Table \ref{demag} presents the list of samples used in magnetic field-employing experiments, together with the description of their shape demagnetization factor calculated using Ref. \cite{Aharoni_1998}. 
\begin{table}[h]
\begin{center}

\begin{ruledtabular}
\caption{Description of shape and demagnetization factors $N$ for TmNiC$_2$ samples used in various experiments.}
\label{demag}

\begin{tabular}{ c c c c c }

 number & experiment type & shape & demagnetization factor $N$ & comment \\ 
 \hline
 1 & magnetization & elongated cuboid & 0.109 & field along the longest edge length \\  
 2 & specific heat & cube & 0.333 & \\
 3 & synchrotron diffraction & plate-like cuboid & 0.093 & field along the longest edge length
\end{tabular}
\end{ruledtabular}

\end{center}
\end{table}

\subsection{Beam heating estimate}

The total power carried by the incoming x-ray beam is expressed by  Eq. \ref{eq_attenuation}:

\begin{equation}
\label{eq_attenuation}
P_{in}=\frac{E_{ph}\Phi}{R_v} 
\end{equation}
where $E_{ph}$ is the single photon energy, $\Phi$ is the photon flux, and $R_v$ is the variable attenuation rate. Considering the beam energy $E_{beam} = 25$ keV, and corresponding intensity $\Phi\approx 3.5\cdot 10^{12}$ photons/s \cite{DESY}, we obtain $P_{in}$ = 28 $\mu$W and 14 $\mu$W, for attenuation rate $R_v$ =  500 and 1000, respectively. These values are close, or smaller than the power carried by the beam in the validating experiment, performed on the same beamline using a TmVO$_4$ sample, which reached the thermal equilibrium at 1.9 K, when cooled down via the conduction mechanism alone, and illuminated by beam power $P$ = 26 $\mu$W \cite{DESY_S}. 
In our experiment, the conduction mechanism transfer is not the only channel, and overall heat evacuation is strongly reinforced by convection with He exchange gas, being even a more effective mechanism. We however focus on the former channel to quantify a reliable worst-case estimate of the maximum temperature rise $\Delta T$. As a reference, we rely on the direct empirical temperature rise determination in the above-mentioned case of TmVO$_4$, and compare the conditions of both experiments and intrinsic properties of relevant materials. The temperature rise $\Delta T$ in our sample is calculated via Eq.~\ref{eq_comparison}, based on Fourier's law of heat conduction:

\begin{equation}
\label{eq_comparison}
\Delta T = \Delta T_{ref}\frac{\kappa_{ref}}{\kappa}\frac{\left( \frac{S}{t} \right)_{ref}}{\frac{S}{t}},
\end{equation}
where $\Delta T_{ref} = 1.54$ K is the temperature rise in TmVO$_4$ - from 360 mK to 1.9 K, $\kappa_{ref}$ and $\kappa$ are heat conductivities, and $\left( \frac{S}{t} \right)_{ref}$ = 5.33 mm, and $\frac{S}{t}$ = 14.37 mm are area/thickness ratios of the reference (TmVO$_4$), and TmNiC$_2$, respectively.
While the various models anticipate $\Delta T$ also to be proportional to $\mu_{att}$ \cite{Kriminski_2003, Mhaisekar2005}, our worst-case estimation does not take this factor into account as for heavy-element based compounds it may show weaker impact on $\Delta T$.
We however note that, $\mu_{att}$ calculated at relevant energies (25 keV for TmNiC2 and 8.650 keV for TmVO$_4$) using Ref.~\cite{XCOM}, and density of both compounds, is approximately 4.7 times higher in TmVO$_4$, than in TmNiC$_2$, which suggests smaller beam heating susceptibility in the latter compound.

We have calculated thermal conductivity of TmNiC$_2$ via the Wiedemann-Franz law, using the measured $c$-axis electrical resistivity \cite{Roman2023}. The resulting $\kappa \approx 0.26$ W m$^{-1}$K$^{-1}$ at 2.9 K, which comprises only the electron term, is close to the experimental result of $\kappa \approx 0.285$ W m$^{-1}$K$^{-1}$ obtained in SmNiC$_2$ at the same temperature \cite{Kim2012}. 
In Eq.~\ref{eq_comparison} we use the calculation result at each temperature and $\kappa_{ref} = 0.115$ W m$^{-1}$K$^{-1}$ for TmVO$_4$ \cite{Daudin_1984}, being the mean value in the range between the initial, and final temperatures (360 mK and 1.9 K, respectively) during the referenced experiment.

Figure~\ref{fig_DeltaT} (a) shows the maximum temperature rise obtained by Eq. \ref{eq_comparison}, which is below 0.3 K at nominal temperature $T= 2.9$ K.  
Furthermore, as depicted in  Fig.~\ref{fig_DeltaT}(b), for all the points corresponding to nominal temperatures below $T_N$, the maximum sample temperature $T+ \Delta T$ remains far away from exceeding the boundaries of the magnetically ordered state.

\renewcommand{\thefigure}{S1}
\begin{figure} [h]
  \includegraphics[angle=0,width=0.90\columnwidth]{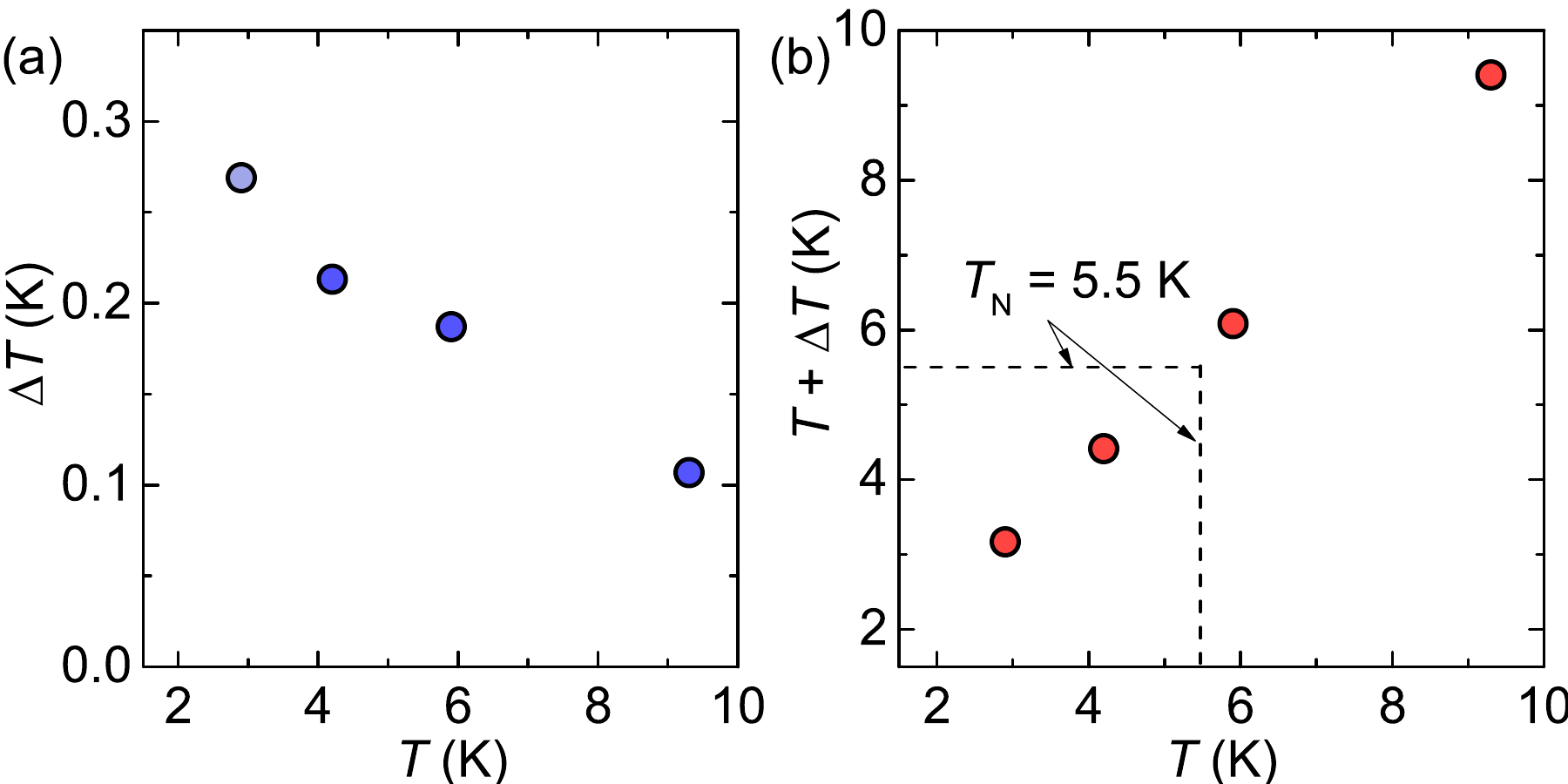}
\caption{\label{fig_DeltaT} (a) Maximum sample temperature increase $\Delta T$, due to beam-heating as obtained using Eq.~\ref{eq_comparison} at all nominal temperatures. 
The point corresponding to $T$ = 2.9 K is highlighted with a brighter color, as the twice large beam attenuation factor applied during the actual measurement at this temperature is not considered in this worst-case estimate via Eq.~\ref{eq_comparison}, and thus, real $\Delta T$, even if taking into account only the conduction mechanism is expected to be lower.  (b) Estimated maximum sample temperature $T+ \Delta T$. Dashed horizontal line shows the N{\'e}el temperature $T_N$ = 5.5 K.}
  \end{figure}

We emphasize, that this most cautious model takes into account only the conduction heat transfer mechanism. Because a more effective mechanism -  convection cooling with He gas cryostream was simultaneously participating in heat evacuation during our experiment, the true temperature rise is lower than provided by our worst case model.

\section{Complementary experimental data}

\subsection{Experimental peak shapes}

Peak shapes and intensities were extracted from 2D detector images
by parsing the SPEC log with a custom script and extracting pixel intensities
with \emph{CBFlib} \cite{cbflib}.
The data was corrected for beam intensity fluctuations and multiplied by the factor for which the incident beam was attenuated.

The $h$-, $k$- and $l$- profiles of $q$ = (-0.5, 5.5, 7.5) and $q$ = (-0.5, 6.5, 8.5) satellite reflections measured at various temperatures including above and below $T_N$ are presented in figures \ref{057QT} and \ref{068QT}, respectively. The broad and split character of the $k$-profiles of the peaks is associated to the twinning and a possible strain, which become more pronounced for reflections indexed with high numbers. The peak profile however does not change significantly in any direction in the reciprocal space either with lowering the temperature or application of magnetic field and entering the subsequent magnetic phases. 

\renewcommand{\thefigure}{S2}
  \begin{figure*} [h!]
  \includegraphics[angle=0,width=0.98\columnwidth]{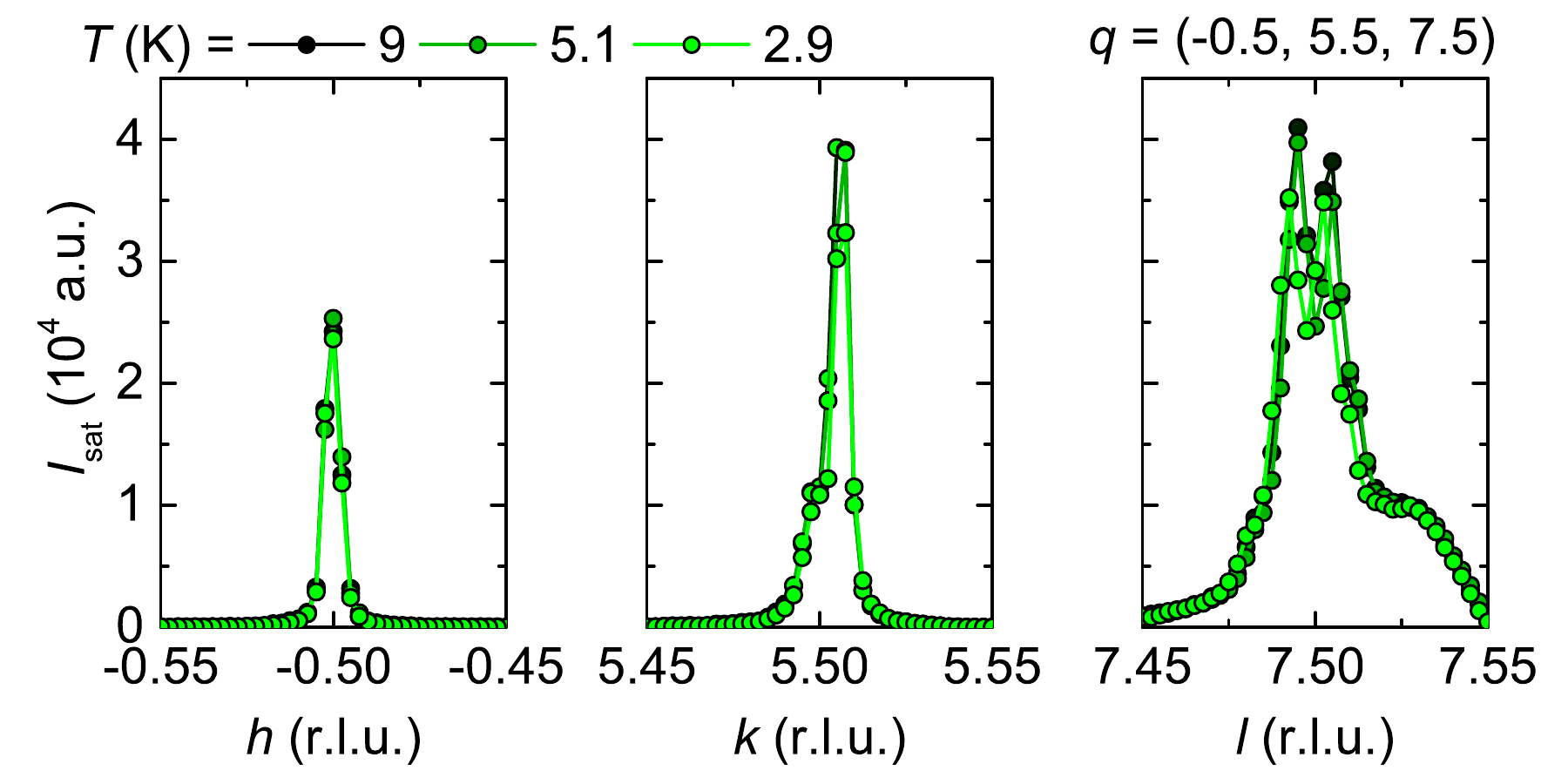}
\caption{\label{057QT} $q$ = (-0.5, 5.5, 7.5) satellite peak profile along three directions of reciprocal lattice at various temperatures. The color legend is situated above the panels.}
  \end{figure*}
  
\renewcommand{\thefigure}{S3}
    \begin{figure*} [h!]
  \includegraphics[angle=0,width=0.98\columnwidth]{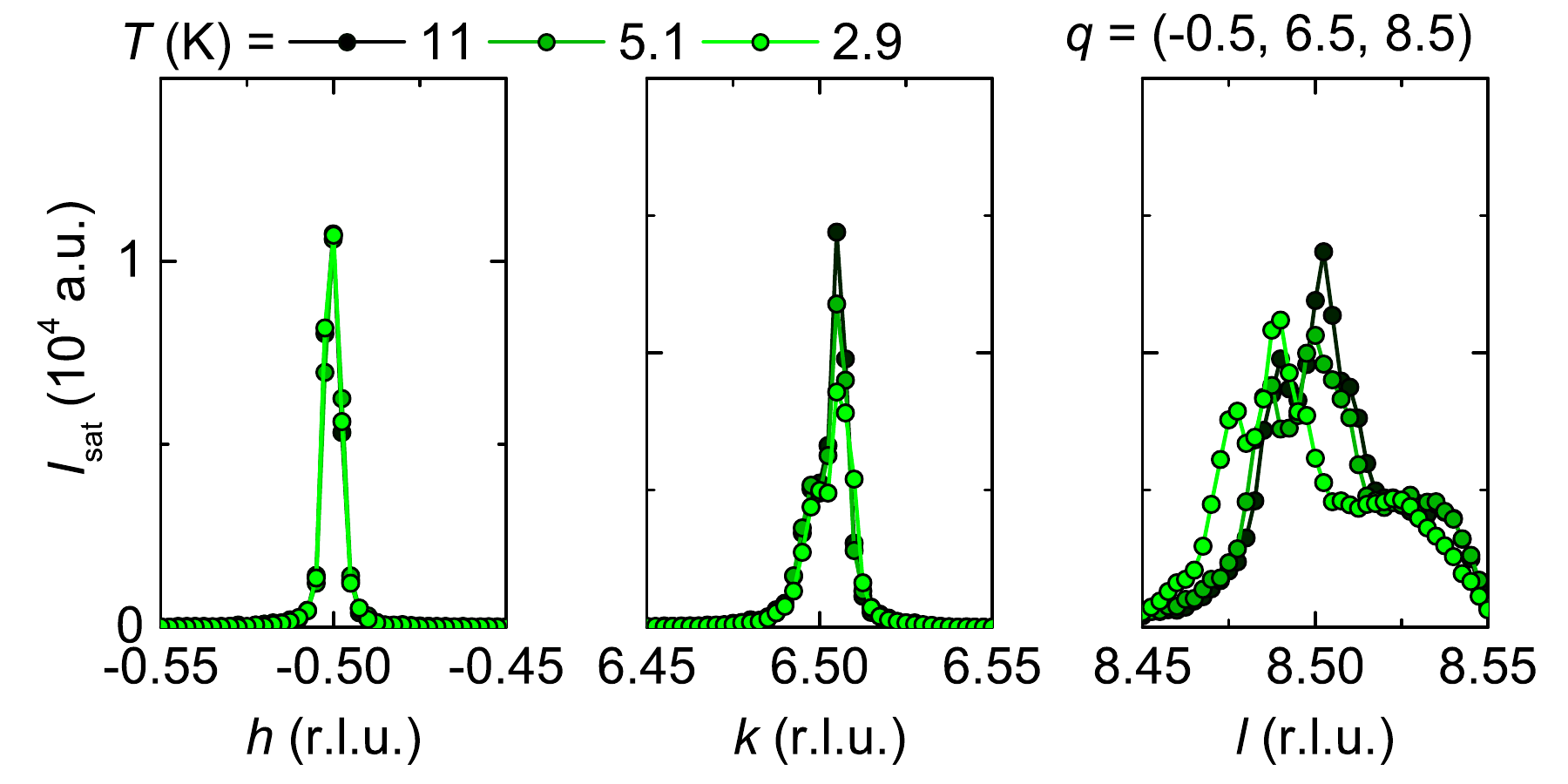}
\caption{\label{068QT} $q$ = (-0.5, 6.5, 8.5) satellite peak profile along three directions of reciprocal lattice at various temperatures. The color legend is situated above the panels.}
  \end{figure*}

The peak profiles measured at $T$ = 2.9 K and various magnetic fields are presented in figures \ref{057QF} [$q$ = (-0.5, 5.5, 7.5)] and \ref{068QF} [$q$ = (-0.5, 6.5, 7.8)].

\renewcommand{\thefigure}{S4}
\begin{figure*} [h!]
  \includegraphics[angle=0,width=0.98\columnwidth]{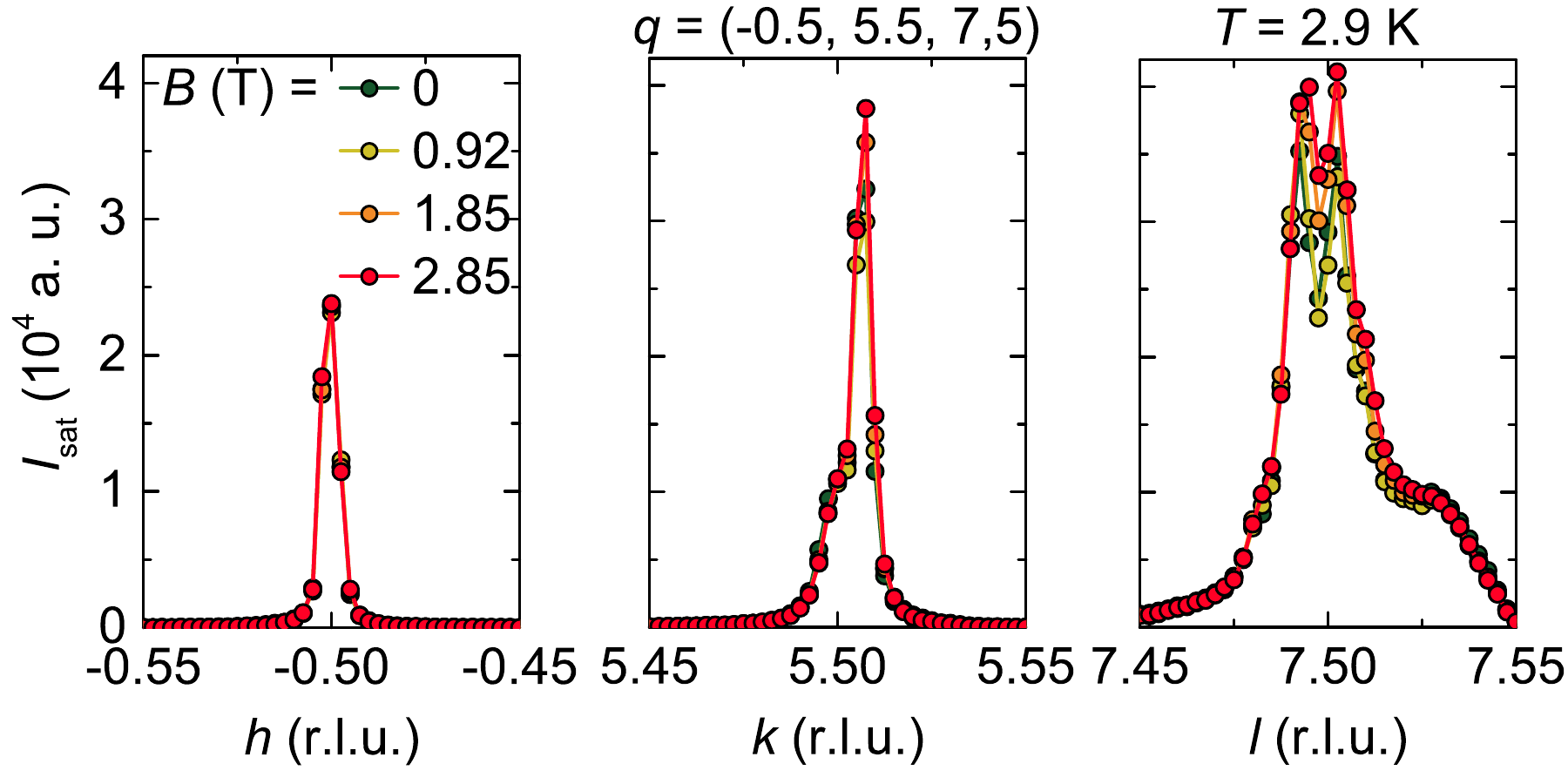}
\caption{\label{057QF} $q$ = (-0.5, 5.5, 7.5) satellite peak profile along three directions of reciprocal lattice at various magnetic fields. The color legend is situated in the left hand panel. All the data presented here has been obtained at $T$ = 2.9 K}
  \end{figure*}

\renewcommand{\thefigure}{S5}
  \begin{figure*} [!htb]
  \includegraphics[angle=0,width=0.98\columnwidth]{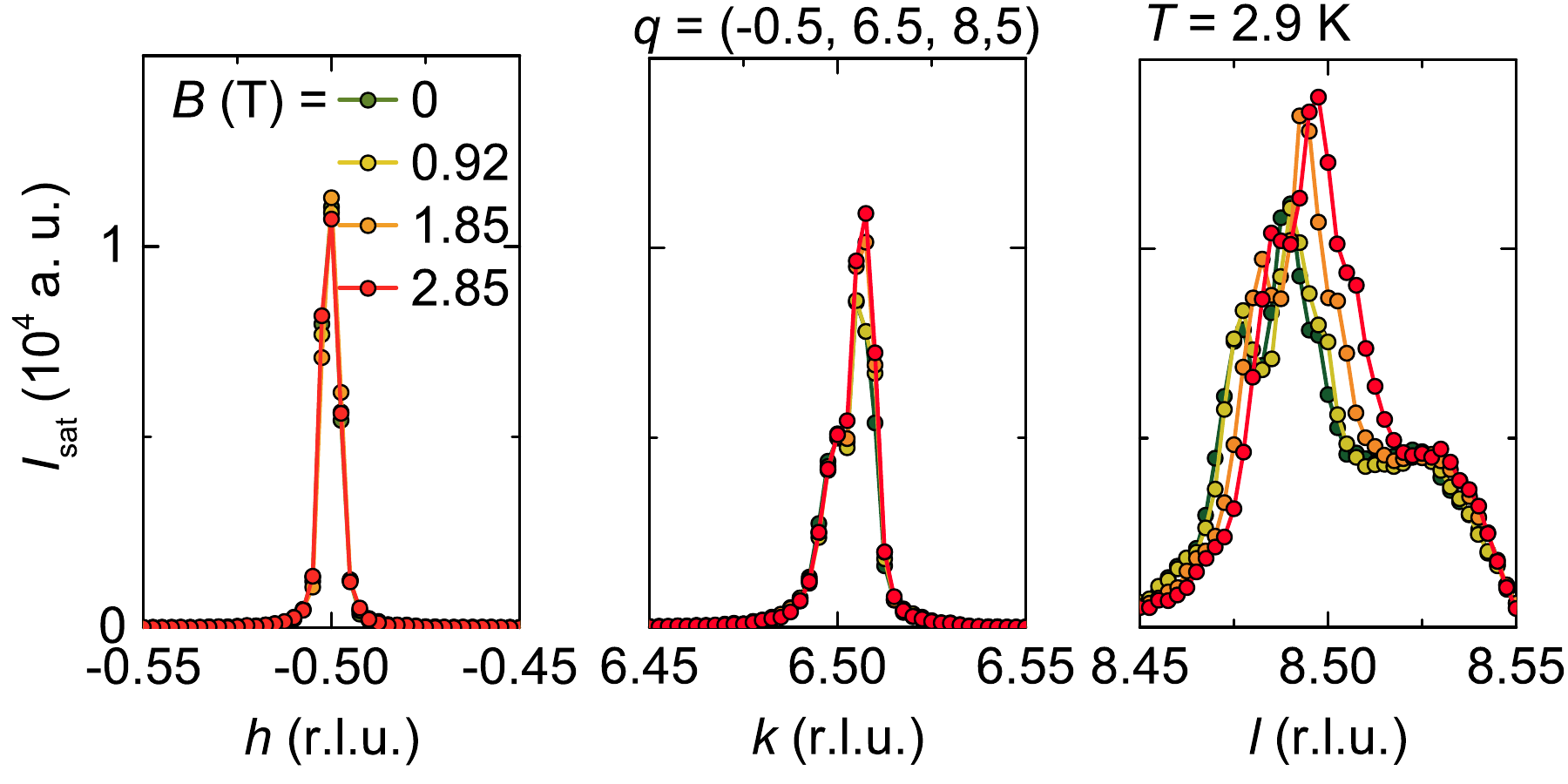}
\caption{\label{068QF} $q$ = (-0.5, 6.5, 8.5) satellite peak profile along three directions of reciprocal lattice at various magnetic fields. The color legend is situated in the left hand panel. All the data presented here has been obtained at $T$ = 2.9 K}
  \end{figure*}
  
The visible peak splitting observed at all temperatures stems from the presence of two twin domains. This is the result of lattice modification and lowering of the crystal symmetry from orthorhombic \textit{Amm2} to monoclinic \textit{Cm} spacegroup, occuring at the onset of CDW transition. Splitting is especially pronounced at high hkl values, corresponding to the peaks available during synchrotron experiment. Figure \ref{hk2} shows the \textit{hk}2 plane measured with laboratory diffractometer. Bragg peak broadening is visible already at $h$ and/or $k$ as low as 2 or 3, even without the synchrotron resolution.

\renewcommand{\thefigure}{S6}
   \begin{figure} [h!]
  \includegraphics[angle=0,width=0.6\columnwidth]{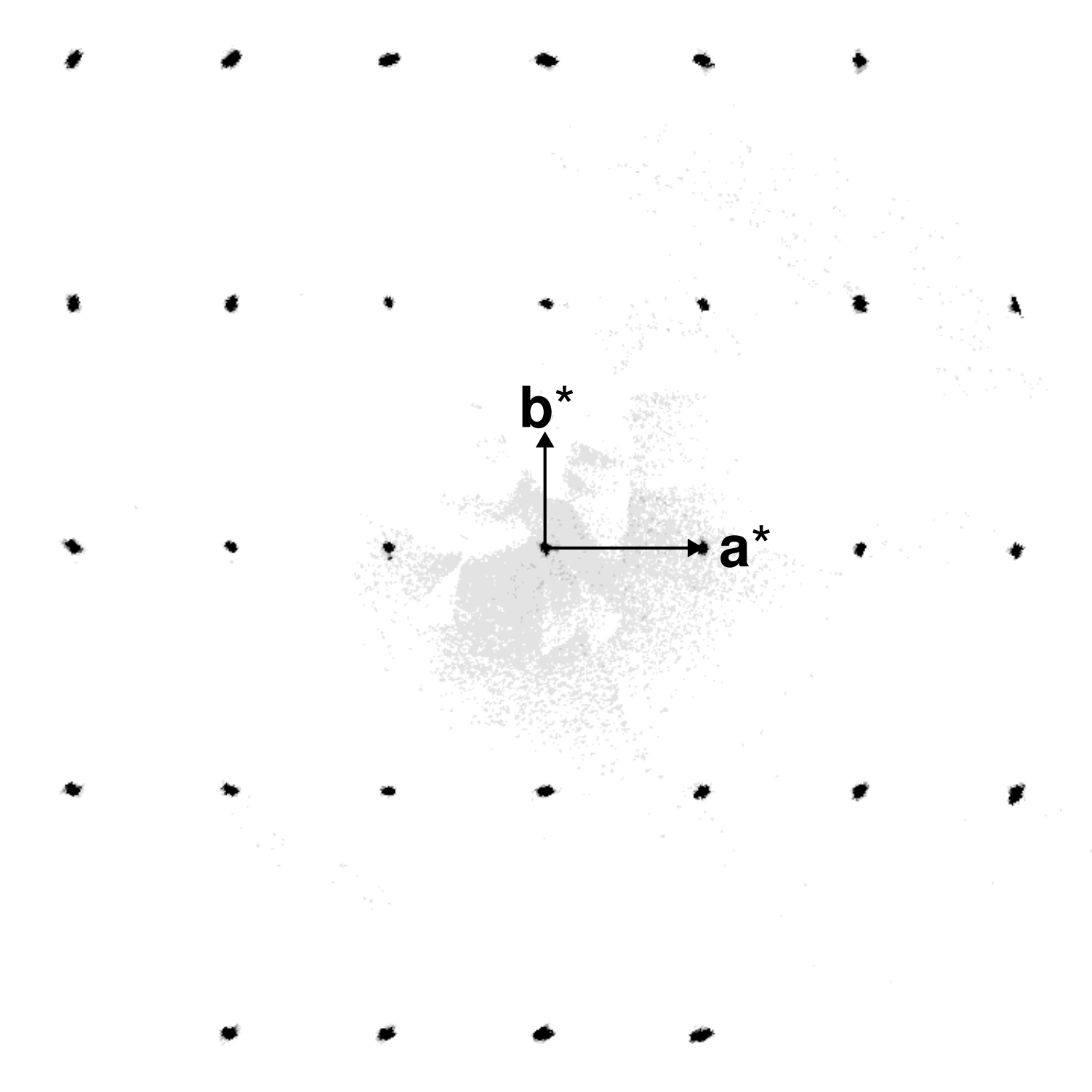}
\caption{\label{hk2} Reconstructed \textit{hk}2 plane of TmNiC$_2$ single crystal. Data was collected with laboratory diffractometer at 100 K.}
  \end{figure}

\section{Additional DFT calculations}
In this section we report additional figures relative to our DFT calculations that complement the results in the main text. In Fig. \ref{bands} we show the band structure of TmNiC$_2$ in the parent and CDW phase in a wider energy range than the main text. In this range, one can observe that the bands in the -4 to 0 eV range are rather flat, and associated with a peaked DOS, corresponding to Ni bands. 

In Fig. \ref{ph}, we report the phonon dispersion calculated directly in the q$_2$-Type CDW phase, as derived in our DFT calculation. The phonon softening seen in the phonon dispersions of the normal phase, at R and between Z and $\mathrm{A_{0}}$, are removed. Hence the q$_2$-Type CDW phase removes the instability seen in the parent phase. Only very a minor softening remains.

\renewcommand{\thefigure}{S7}
   \begin{figure} [h!]
  \includegraphics[angle=0,width=0.7\columnwidth]{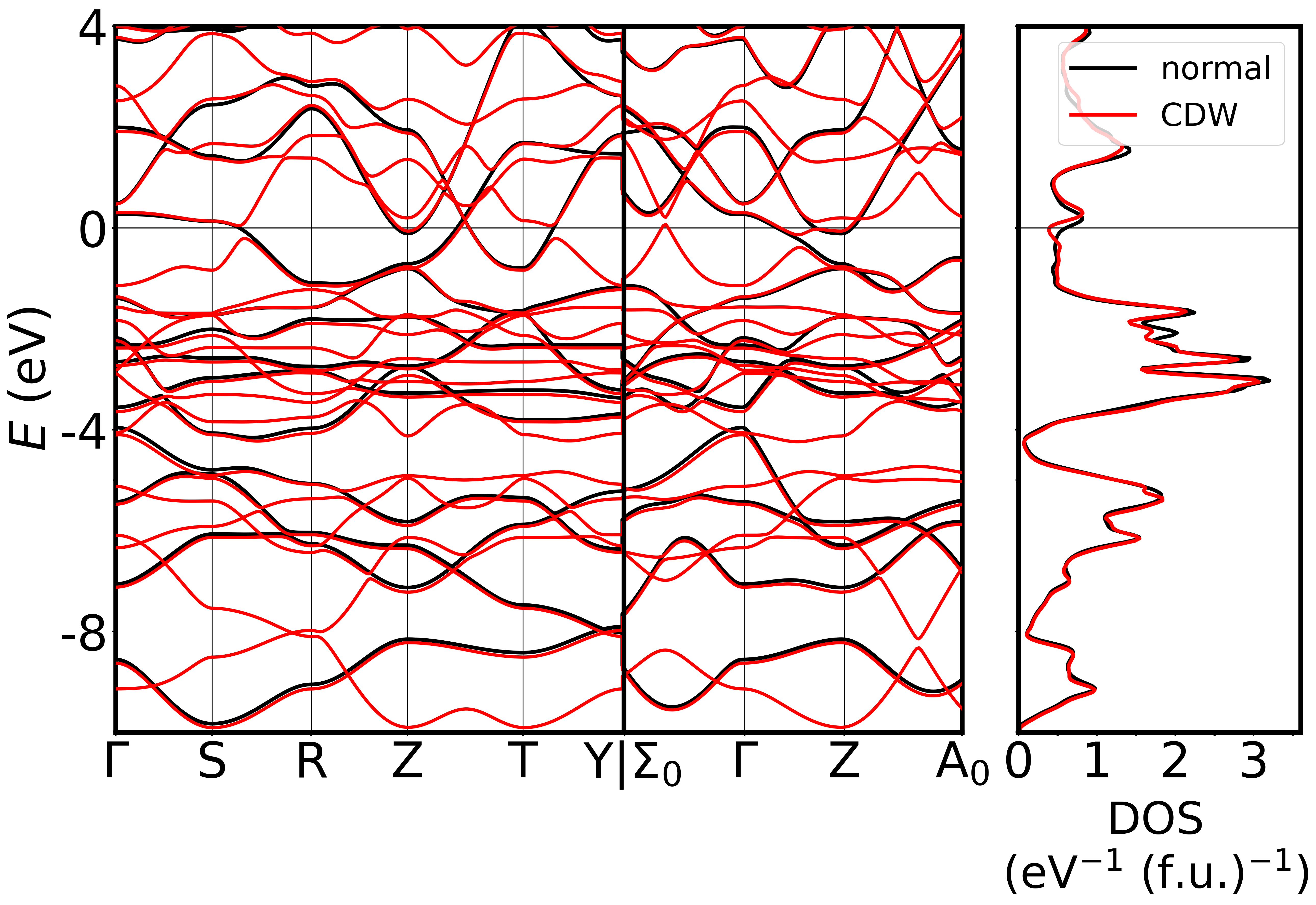}
\caption{Electronic band structure and DOS of $\mathrm{TmNiC}_{2}$ in a wide energy range. The DFT bands for the normal and q$_2$-Type CDW are shown as black and red lines, respectively.}
\label{bands}
  \end{figure}

\renewcommand{\thefigure}{S8}
   \begin{figure} [h!]
  \includegraphics[angle=0,width=0.7\columnwidth]{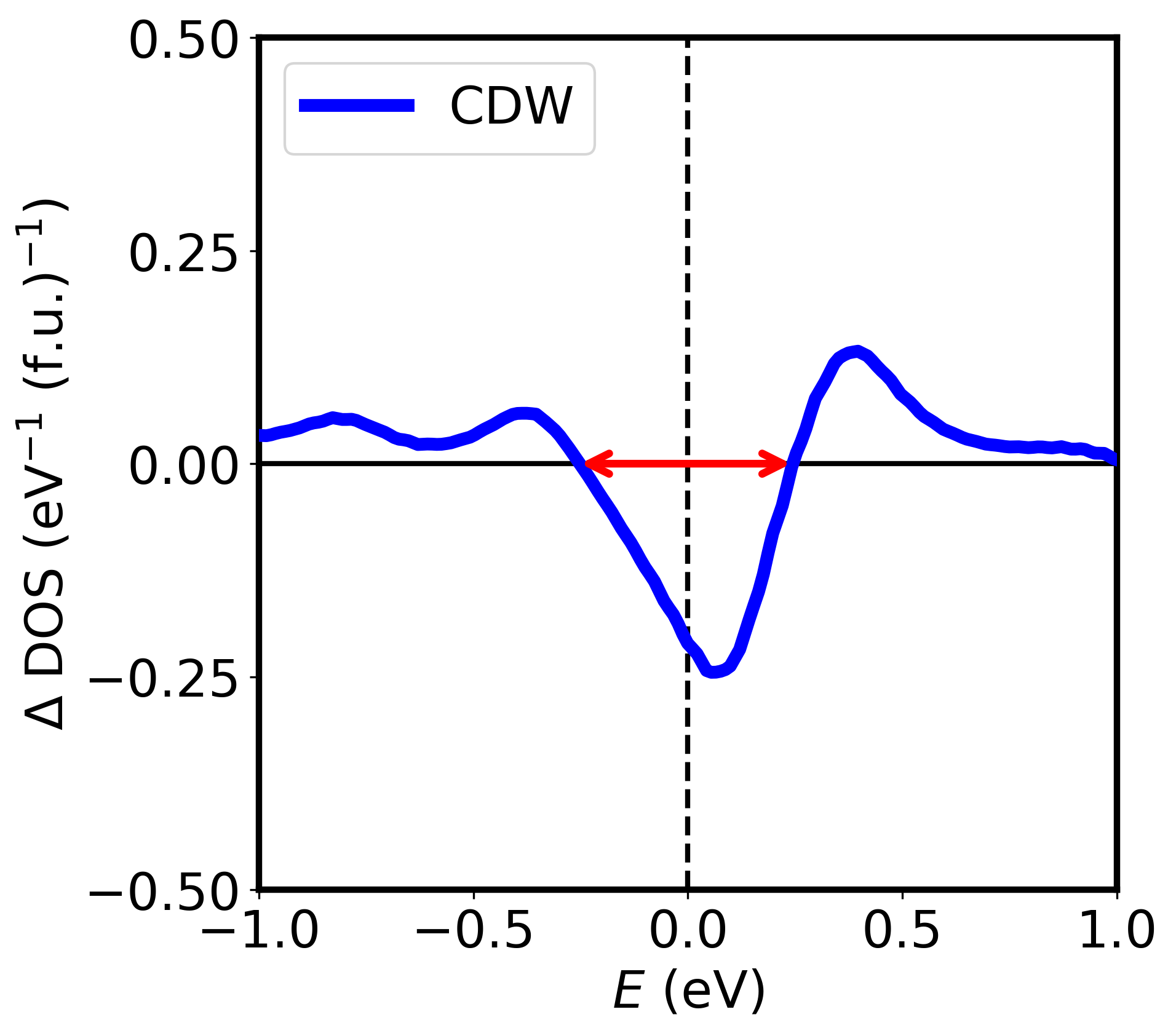}
\caption{Difference between the DOS of the $q_{2}$-type CDW and the parent phase around the Fermi energy. The red arrow indicates the width of the CDW gap.}
\label{dosdifference}
  \end{figure}

\renewcommand{\thefigure}{S9}
  \begin{figure} [h!]
  \includegraphics[angle=0,width=0.6\columnwidth]{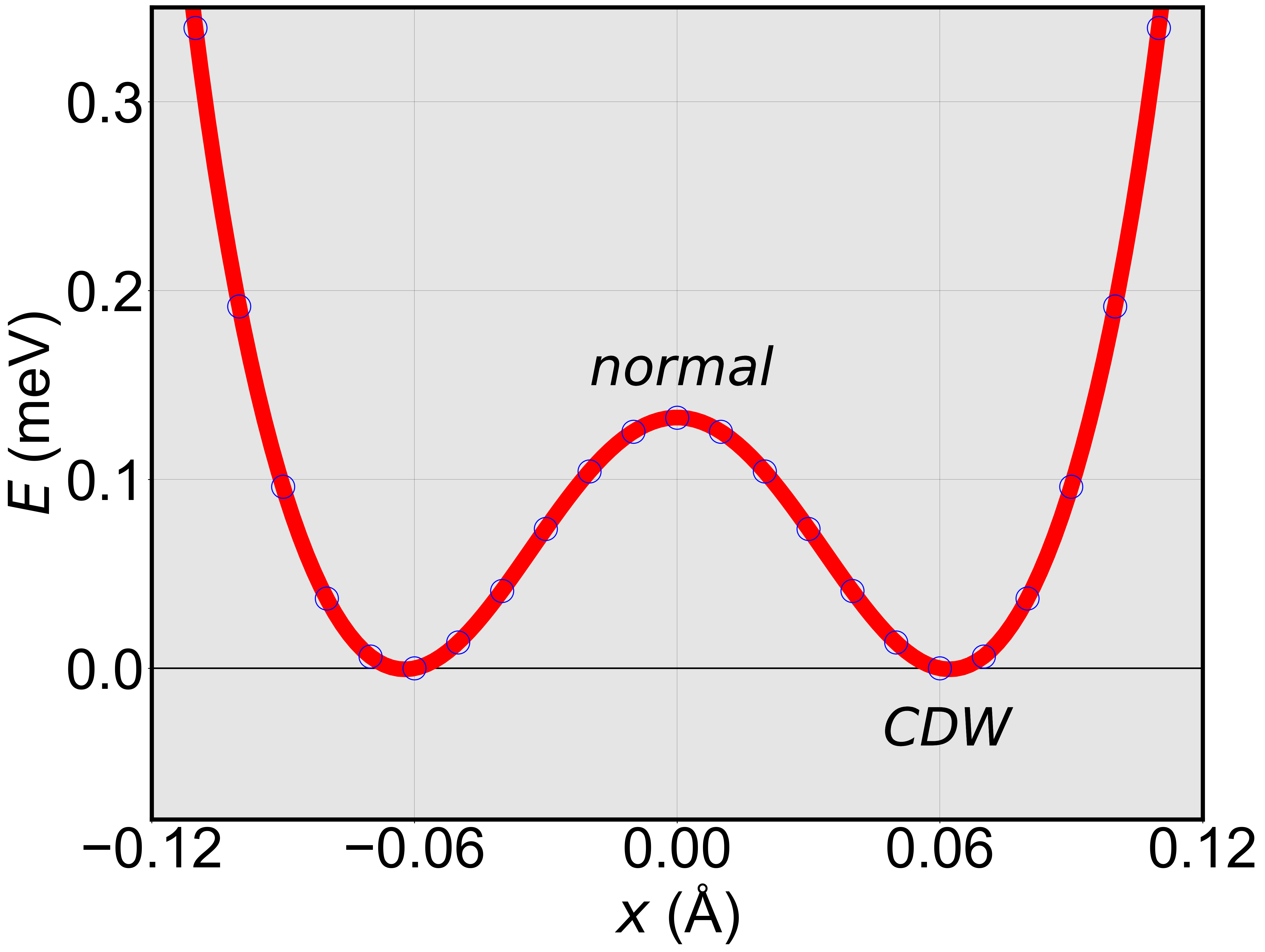}
\caption{Potential energy well with a displacement along the soft eigenvector for $q_{2c}$. $x$ represents the collective displacement of the unstable phonon in the parent phase.}
\label{bands}
  \end{figure}

\renewcommand{\thefigure}{S10}
   \begin{figure} [h!]
  \includegraphics[angle=0,width=0.7\columnwidth]{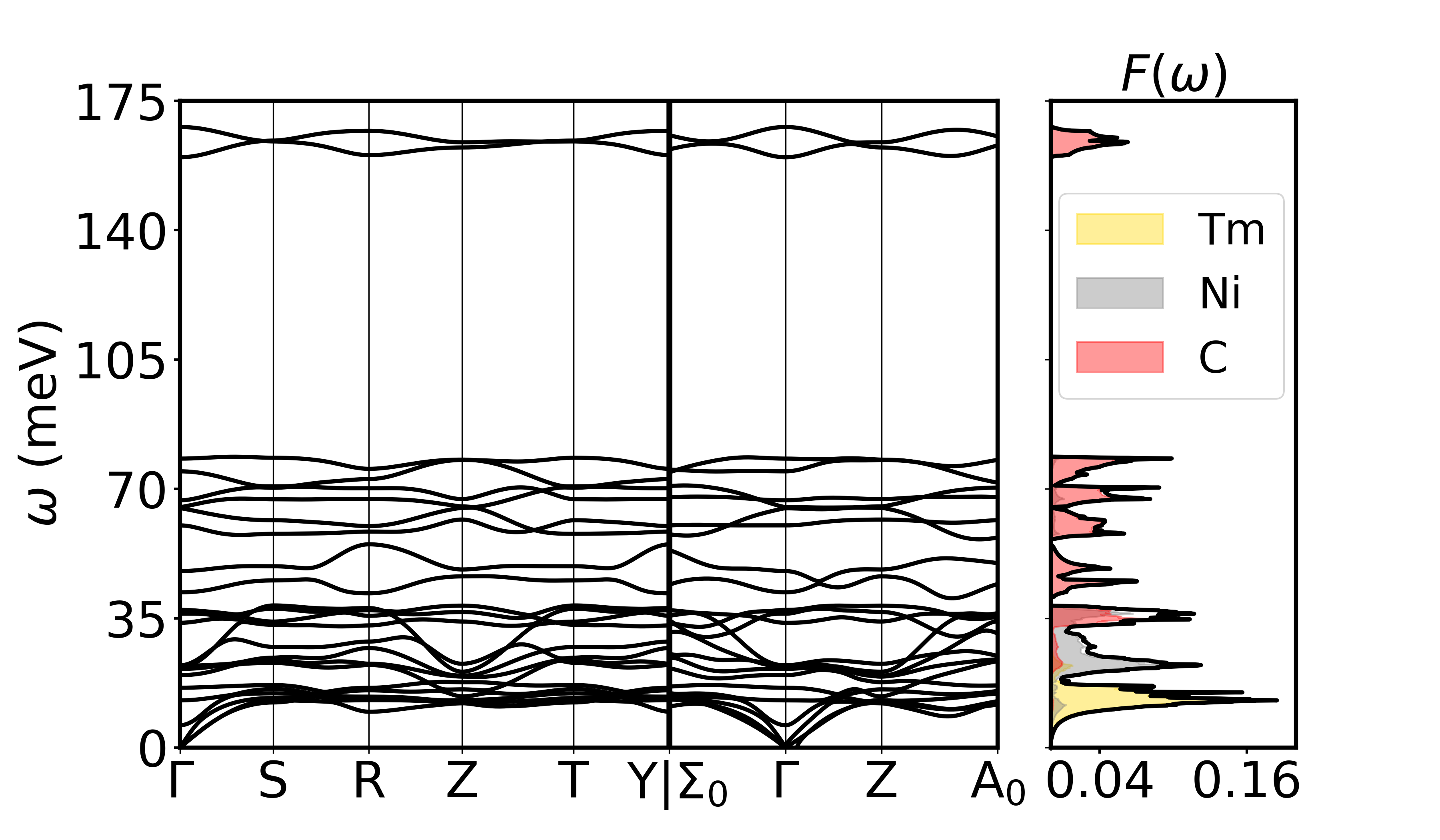}
\caption{Phonon dispersions of the $q_{2c}$-type CDW phase of $\mathrm{TmNiC_{2}}$, along with atom-projected phonon density of states. Projection onto Tm, Ni, and C are shown as yellow, gray, and red filled lines.}
\label{ph}
  \end{figure}

\clearpage

% \bibliography{HE_bibliography}
%